\documentclass{iopart}

\usepackage{iopams}
\usepackage{graphicx}

\newcommand{\der}[2]{\frac{\partial #1}{\partial #2}}
\newcommand{\dert}[2]{{\partial #1}/{\partial #2}}
\newcommand{\w}[1]{\boldsymbol{#1}}
\newcommand{\Lie}[1]{\boldsymbol{\mathcal{L}}_{\w{#1}}\,}
\newcommand{\Liec}[1]{{\mathcal{L}}_{\w{#1}}\,}
\newcommand{\wpar}{\w{\partial}}

\newcommand{\be}{\begin{equation}}
\newcommand{\ee}{\end{equation}}
\newcommand{\bea}{\begin{eqnarray}}
\newcommand{\eea}{\end{eqnarray}}

\newcommand{\R}{\mathbb{R}}
\newcommand{\Sp}{\mathcal{S}}
\newcommand{\Df}{\mathcal{D}}
\newcommand{\wDf}{\w{\Df}}
\newcommand{\gm}{\gamma}
\newcommand{\wgm}{\w{\gm}}
\newcommand{\wD}{\w{D}}
\newcommand{\tgm}{{\tilde\gm}}
\newcommand{\wtgm}{\w{\tgm}}
\newcommand{\tD}{{\tilde D}}
\newcommand{\wtD}{\w{\tD}}
\newcommand{\tA}{{\tilde A}}
\newcommand{\hA}{{\hat A}}

\newenvironment{remark}%
{\begin{description} \item[\emph{Remark :}]\em}%
{\end{description}}

\begin{document}

\title{Construction of initial data for 3+1 numerical 
relativity}

\author{Eric Gourgoulhon}
\address{Laboratoire Univers et Th\'eories, 
UMR 8102 du C.N.R.S., Observatoire de Paris, 
Universit\'e Paris 7 - Denis Diderot, F-92195 Meudon Cedex,  France}
\ead{eric.gourgoulhon@obspm.fr}

\begin{abstract}
This lecture is devoted to the problem of computing initial data for the Cauchy
problem of 3+1 general relativity. The main task is to solve the constraint equations. 
The conformal technique, introduced by Lichnerowicz and enhanced by York, is presented. 
Two standard methods, the conformal transverse-traceless one and the conformal 
thin sandwich, are discussed and illustrated by some simple examples.
Finally a short review regarding initial data for binary systems
(black holes and neutron stars) is given.  
\end{abstract}

\submitto{Journal of Physics: Conference Series, for 
the Proceedings of the VII Mexican School on Gravitation and Mathematical Physics,
held in Playa del Carmen, Quintana Roo, Mexico
(November 26 - December 2, 2006)}



\section{Introduction}

The 3+1 formalism is the basis of most modern numerical relativity
and has lead, along with alternative approaches 
\cite{Preto05b}, to the recent successes in the 
binary black hole merger problem \cite{BakerCCKV06a,BakerCCKV06b,VanMeBKC06,CampaLMZ06,CampaLZ06a,CampaLZ06b,CampaLZ06c}
(see \cite{Campa07,Lagun07,Shoem07} for a review).
Thanks to the 3+1 formalism, 
the resolution of Einstein equation amounts to solving a Cauchy
problem, namely to evolve ``forward in time'' some initial data.
However this is a Cauchy problem with constraints. This makes the set up
of initial data a non trivial task, because these data must fulfill
the constraints. In this lecture, we present the most wide spread methods 
to deal with this problem. Notice that we do not 
discuss the numerical techniques employed to solve the constraints
(see e.g. Choptuik's lecture for finite differences 
\cite{Chopt07} and 
Grandcl\'ement and Novak's review for spectral methods \cite{GrandN07}). 

Standard reviews about the initial data problem are the articles by
York \cite{York79} and Choquet-Bruhat and York 
\cite{ChoquY80}. Recent reviews are the articles by
Cook \cite{Cook00}, Pfeiffer \cite{Pfeif04}
and Bartnik and Isenberg \cite{BartnI04}.

\section{The initial data problem}

\subsection{3+1 decomposition of Einstein equation}

In this lecture, we consider a spacetime $(\mathcal{M},\w{g})$, where $\mathcal{M}$ is a four-dimensional smooth manifold and 
$\w{g}$ a Lorentzian metric on $\mathcal{M}$. We assume that 
$(\mathcal{M},\w{g})$ is \emph{globally hyperbolic}, i.e. that $\mathcal{M}$ can be 
foliated by a family $(\Sigma_t)_{t\in\R}$ of spacelike hypersurfaces.
We denote by $\wgm$ the (Riemannian) metric induced by $\w{g}$ on each hypersurface
$\Sigma_t$ and $\w{K}$ the extrinsic curvature of $\Sigma_t$, with the same
sign convention as that used in the numerical relativity community, i.e. for any
pair of vector fields $(\w{u},\w{v})$ tangent to $\Sigma_t$,
$\w{g}(\w{u},\w{\nabla}_{\w{v}}\w{n})=-\w{K}(\w{u},\w{v})$, where $\w{n}$ is the
future directed unit normal to $\Sigma_t$ and $\w{\nabla}$ is the Levi-Civita connection
associated with $\w{g}$. 

The 3+1 decomposition of Einstein equation with respect to
the foliation $(\Sigma_t)_{t\in\R}$ leads to three sets of equations:
(i) the \emph{evolution equations} of the Cauchy problem (full
projection of Einstein equation onto $\Sigma_t$), (ii) the \emph{Hamiltonian constraint}
(full projection of Einstein equation along the normal $\w{n}$),
(iii) the \emph{momentum constraint} (mixed projection: once onto $\Sigma_t$, once
along $\w{n}$). The latter two sets of equations do not contain any second derivative
of the metric with respect to $t$. They are written\footnote{we are using the standard
convention for indices, namely Greek indices run in $\{0,1,2,3\}$, whereas Latin
ones run in $\{1,2,3\}$}
\bea
 & &  R + K^2 - K_{ij} K^{ij} = 16\pi E \quad \mbox{(Hamiltonian constraint)}, \label{e:ini:Ham_constr}\\
 & &  D^j K_{ij} - D_i K = 8\pi  p_i 
	\quad \mbox{(momentum constraint)} , \label{e:ini:mom_constr}
\eea
where $R$ is the Ricci scalar (also called \emph{scalar curvature})
associated with the 3-metric $\wgm$, $K$ is the trace
of $\w{K}$ with respect to $\wgm$: $K=\gm^{ij} K_{ij}$, $\w{D}$ stands for the Levi-Civita connection associated
with the 3-metric $\wgm$, and $E$ and $p_i$ are respectively the energy density and
linear momentum of matter, both measured by the observer of 4-velocity $\w{n}$
\emph{(Eulerian observer)}. In terms of the matter energy-momentum tensor $\w{T}$
they are expressed as 
\be
	E = T_{\mu\nu} n^\mu n^\nu
	\qquad\mbox{and}\qquad
	p_i = - T_{\mu\nu} n^\mu \gm^\nu_{\ \, i} . 
\ee
Notice that Eqs.~(\ref{e:ini:Ham_constr})-(\ref{e:ini:mom_constr}) involve 
a single hypersurface $\Sigma_0$, not a foliation $\left(\Sigma_t\right)_{t\in \R}$. 
In particular, neither the lapse function nor the shift vector appear in these equations. 

\subsection{Constructing initial data} \label{s:ini:idp}

In order to get valid initial data for the Cauchy problem, 
one must find solutions to the constraints (\ref{e:ini:Ham_constr}) and (\ref{e:ini:mom_constr}).
Actually one may distinguish two problems:
\begin{itemize}
\item \emph{The mathematical problem:} given some hypersurface $\Sigma_0$, 
find a Riemannian metric $\wgm$, a symmetric bilinear form $\w{K}$
and some matter distribution $(E,\w{p})$ on $\Sigma_0$
such that the Hamiltonian constraint
(\ref{e:ini:Ham_constr}) and the momentum constraint (\ref{e:ini:mom_constr})
are satisfied.
In addition, the matter distribution $(E,\w{p})$ may have some constraints
from its own. We shall not discuss them here.
\item \emph{The astrophysical problem:} make sure that the solution
to the constraint equations has something to do with the physical system 
that one wish to study. 
\end{itemize}
Facing the constraint equations (\ref{e:ini:Ham_constr}) and (\ref{e:ini:mom_constr}), 
a naive way to proceed would be to choose freely the metric $\wgm$, thereby
fixing the connection $\w{D}$ and the scalar curvature $R$, and to solve 
Eqs.~(\ref{e:ini:Ham_constr})-(\ref{e:ini:mom_constr}) for $\w{K}$.
Indeed, for fixed $\wgm$, $E$, and $\w{p}$, Eqs.~(\ref{e:ini:Ham_constr})-(\ref{e:ini:mom_constr}) form a quasi-linear system 
of first order for the components $K_{ij}$. 
However, as discussed by Choquet-Bruhat \cite{Foure56}, this 
approach is not satisfactory
because we have only four equations for six unknowns $K_{ij}$ and there is
no natural prescription for choosing arbitrarily two among the six components
$K_{ij}$. 

In 1944, Lichnerowicz \cite{Lichn44} has shown that a
much more satisfactory split of the initial data $(\wgm,\w{K})$
between freely choosable parts and parts obtained by solving Eqs.~(\ref{e:ini:Ham_constr})-(\ref{e:ini:mom_constr})
is provided by a conformal decomposition of the metric $\wgm$. 
Lichnerowicz method has been extended by Choquet-Bruhat (1956, 1971) 
\cite{Foure56,Choqu71},
by York and \'O Murchadha (1972, 1974, 1979)
\cite{York72b,York73,OMurcY74,York79} and more recently by York and Pfeiffer
(1999, 2003) \cite{York99,PfeifY03}. 
Actually, conformal decompositions are by far the 
most widely spread techniques to get initial data for the 3+1
Cauchy problem. 
Alternative methods exist, such as the quasi-spherical ansatz introduced by
Bartnik in 1993 \cite{Bartn93} or a procedure developed by Corvino (2000)
\cite{Corvi00} and by Isenberg,
Mazzeo and Pollack (2002) \cite{IsenbMP02} for gluing together known solutions of
the constraints, thereby producing new ones. 
Here we shall limit ourselves to the conformal methods. 

\subsection{Conformal decomposition of the constraints}

In the conformal approach initiated by Lichnerowicz \cite{Lichn44}, 
one introduces a \emph{conformal metric} $\wtgm$ and a \emph{conformal factor} $\Psi$
such that the (physical) metric $\wgm$ induced by the spacetime metric on the hypersurface
$\Sigma_t$ is
\be \label{e:ini:gm_Psi_tgm}
	\gm_{ij} = \Psi^4 \tgm_{ij} . 
\ee
We could fix some degree of freedom by demanding that $\det\tgm_{ij}=1$. This 
would imply $\Psi = (\det\gm_{ij})^{1/12}$. However, in this case $\wtgm$ and 
$\Psi$ would be tensor densities.
Moreover the condition $\det\tgm_{ij}=1$ has a meaning only for Cartesian-like coordinates.
In order to deal with tensor fields and to allow for any type of coordinates, 
we proceed differently and introduce a \emph{background} Riemannian metric $\w{f}$ on $\Sigma_t$.
If the topology of $\Sigma_t$ allows it, we shall demand that $\w{f}$ is flat.
Then we replace the condition $\det\tgm_{ij}=1$ by $\det\tgm_{ij}=\det f_{ij}$.
This fixes
\be
	\Psi = \left( \frac{\det\gm_{ij}}{\det f_{ij}} \right) ^{1/12} .
\ee
$\Psi$ is then a genuine scalar field on $\Sigma_t$ (as a quotient of two determinants).
Consequently $\wtgm$ is a tensor field and not a tensor density.

Associated with the above conformal transformation, there are two decompositions
of the traceless part $A_{ij}$ of the extrinsic curvature, the latter being defined by
\be \label{e:K_A_K}
	K_{ij} =: A_{ij} + \frac{1}{3} K \gm_{ij} .
\ee
These two decompositions are 
\bea
	A^{ij} & =: & \Psi^{-10} \hA^{ij} , \label{e:def_hA} \\
	A^{ij} & =: & \Psi^{-4} \tA^{ij} .  \label{e:def_tA}
\eea
The choice $-10$ for the exponent of $\Psi$ in Eq.~(\ref{e:def_hA}) is
motivated by the following identity, valid for any symmetric and traceless tensor field,
\be \label{e:cfd:divA_Psi10}
	D_j A^{ij} = \Psi^{-10} \tD_j \left( \Psi^{10} A^{ij} \right) ,
\ee
where $\tD_j$ denotes the covariant derivative associated with the conformal
metric $\wtgm$. This choice is well adapted to the momentum constraint, because the
latter involves the divergence of $\w{K}$.
The alternative choice, i.e. Eq.~(\ref{e:def_tA}), is motivated by time evolution
considerations, as we shall discuss below. For the time being, we limit
ourselves to the decomposition (\ref{e:def_hA}), having in mind to simplify the
writing of the momentum constraint. 

By means of the decompositions (\ref{e:ini:gm_Psi_tgm}), (\ref{e:K_A_K})
and (\ref{e:def_hA}), the Hamiltonian constraint (\ref{e:ini:Ham_constr}) and
the momentum constraint (\ref{e:ini:mom_constr}) are rewritten as 
(see Ref.~\cite{Gourg07a} for details)
\bea
	& & 	 
	\tD_i \tD^i \Psi -\frac{1}{8} {\tilde R} \Psi
	+ \frac{1}{8} \hA_{ij} \hA^{ij} \, \Psi^{-7}
	+ 2\pi {\tilde E} \Psi^{-3} - \frac{1}{12} K^2 \Psi^5 = 0  , 
	 \label{e:ini:Ham_conf} \\
	& & 
	 \tD_j \hA^{ij} - \frac{2}{3} \Psi^6 \tD^i K = 8\pi {\tilde p}^i  , 
		\label{e:ini:mom_conf} 
\eea
where $\tilde R$ is the Ricci scalar associated with the conformal metric $\wtgm$
and we have introduced the rescaled matter quantities
\be \label{e:ini:tE_def}
	{\tilde E} := \Psi^8 E \qquad \mbox{and}\qquad
	{\tilde p}^i := \Psi^{10} p^i .
\ee
Equation~(\ref{e:ini:Ham_conf}) is known as \emph{Lichnerowicz equation},
or sometimes \emph{Lichnerowicz-York equation}.
The definition of ${\tilde p}^i$ is such that there is no $\Psi$ factor in the
right-hand side of Eq.~(\ref{e:ini:mom_conf}).
On the contrary the power $8$ in the definition of $\tilde E$ is not the 
only possible choice. As we shall see in \S~\ref{s:ini:Lichne}, 
it is chosen (i) to guarantee a negative power of $\Psi$ in the $\tilde E$ term in Eq.~(\ref{e:ini:Ham_conf}), resulting in some uniqueness property of the solution
and (ii) to allow for an easy implementation of the dominant energy condition. 

\section{Conformal transverse-traceless method} \label{s:ini:CTT}

\subsection{Longitudinal/transverse decomposition of $\hA^{ij}$} \label{s:ini:long_trans}

In order to solve the system (\ref{e:ini:Ham_conf})-(\ref{e:ini:mom_conf}),
York (1973,1979) \cite{York73,York74,York79} has decomposed
$\hA^{ij}$ into a longitudinal part and a transverse one, setting
\be \label{e:ini:decomp_hA}
	 \hA^{ij} = (\tilde L X)^{ij} + \hA^{ij}_{\rm TT} , 
\ee
where $\hA^{ij}_{\rm TT}$ is both traceless and transverse (i.e. divergence-free)
with respect to the metric $\wtgm$:
\be \label{e:ini:hA_TT}
	\tgm_{ij} \hA^{ij}_{\rm TT} = 0
	\qquad \mbox{and} \qquad
	\tD_j \hA^{ij}_{\rm TT} = 0 ,
\ee
and $(\tilde L X)^{ij}$ is the \emph{conformal Killing operator} associated
with the metric $\wtgm$ and acting on the vector field $\w{X}$:
\be \label{e:ini:conf_Killing_def}
	 (\tilde L X)^{ij} := \tD^i X^j + \tD^j X^i 
			- \frac{2}{3} \tD_k X^k \, \tgm^{ij}  . 
\ee
$(\tilde L X)^{ij}$ is by construction traceless:
\be
	\tgm_{ij} (\tilde L X)^{ij} = 0  
\ee
(it must be so
because in Eq.~(\ref{e:ini:decomp_hA}) both $\hA^{ij}$ and $\hA^{ij}_{\rm TT}$
are traceless).
The kernel of $\w{\tilde L}$ is made of the \emph{conformal Killing vectors}
of the metric $\wtgm$, i.e. the generators of the conformal 
isometries (see e.g. Ref.~\cite{Gourg07a} for more details). 
The symmetric tensor $(\tilde L X)^{ij}$ 
is called the \emph{longitudinal part} of $\hA^{ij}$, whereas $\hA^{ij}_{\rm TT}$
is called the \emph{transverse part}. 

Given $\hA^{ij}$, the vector $\w{X}$ is determined by taking the divergence 
of Eq.~(\ref{e:ini:decomp_hA}): taking into account property (\ref{e:ini:hA_TT}), 
we get
\be \label{e:ini:tDLX_divA}
	\tD_j (\tilde L X)^{ij} = \tD_j \hA^{ij} .
\ee
The second order operator $\tD_j (\tilde L X)^{ij}$ acting on the vector $\w{X}$
is the \emph{conformal vector Laplacian} $\w{\tilde\Delta}_L$:
\be
   \tilde\Delta_L \, X^i := \tD_j (\tilde L X)^{ij}
	= \tD_j \tD^j X^i + \frac{1}{3} \tD^i \tD_j X^j
	+ {\tilde R}^i_{\ \, j} X^j , 
\ee
where the second equality follows from the Ricci identity applied to the
connection $\wtD$, ${\tilde R}_{ij}$ being the associated Ricci tensor.
The operator $\w{\tilde\Delta}_L$ is elliptic and its kernel is, 
in practice, reduced to the conformal Killing vectors of $\wtgm$, if any. We rewrite Eq.~(\ref{e:ini:tDLX_divA}) as
\be \label{e:ini:DeltaLX_divA}
	\tilde\Delta_L \, X^i = \tD_j \hA^{ij} .
\ee
The existence and uniqueness of the longitudinal/transverse decomposition
(\ref{e:ini:decomp_hA}) depend on the existence and uniqueness of solutions
$\w{X}$ to Eq.~(\ref{e:ini:DeltaLX_divA}).
We shall consider two cases:  
\begin{itemize}
\item $\Sigma_0$ is a \emph{closed manifold}, i.e. is compact without boundary;
\item $(\Sigma_0,\wgm)$ is an \emph{asymptotically flat manifold}, i.e. is such
that the background metric $\w{f}$ is flat (except possibly on a compact sub-domain 
$\mathcal{B}$ of $\Sigma_t$) and 
there exists a coordinate system $(x^i)=(x,y,z)$ on $\Sigma_t$ such that
outside $\mathcal{B}$, the components of $\w{f}$ are $f_{ij} = \mathrm{diag}(1,1,1)$
(``Cartesian-type coordinates'') and the variable $r:=\sqrt{x^2+y^2+z^2}$ can
take arbitrarily large values on $\Sigma_t$; then when $r\rightarrow +\infty$, the components of $\w{\gm}$ and $\w{K}$ with respect to the coordinates $(x^i)$ satisfy
\bea
	& & \gm_{ij} = f_{ij} + O(r^{-1})
	\qquad \mbox{and} \qquad
	 \der{\gm_{ij}}{x^k} = O(r^{-2}),	\label{e:glob:aflat12}\\
	& & K_{ij} = O(r^{-2})
	\qquad \mbox{and} \qquad
	\der{K_{ij}}{x^k} = O(r^{-3}).\label{e:glob:aflat34}
\eea
\end{itemize}
In the case of a closed manifold, one can show (see Appendix~B of Ref.~\cite{Gourg07a} for details)
that solutions
to Eq.~(\ref{e:ini:DeltaLX_divA}) exist provided that the source $\tD_j \hA^{ij}$
is orthogonal to all conformal Killing vectors of $\wtgm$, in the sense
that
\be 
	\forall \w{C}\in\mathrm{ker}\, \w{\tilde L},\quad 
	 \int_{\Sigma}  \tgm_{ij}  C^i \tD_k \hA^{jk} \sqrt{\tgm} \, d^3 x = 0  . 
\ee
But the above property is easy to verify:
using the fact that the source is a pure divergence and that $\Sigma_0$
is closed, we may integrate the left-hand side by parts and get, for any vector field $\w{C}$,
\be
	\int_{\Sigma_0}  \tgm_{ij}  C^i \, \tD_k \hA^{jk} \sqrt{\tgm} \, d^3 x = 
	- \frac{1}{2}
	\int_{\Sigma_0}  \tgm_{ij}  \tgm_{kl}  (\tilde L C)^{ik} \hA^{jl}
		\sqrt{\tgm} \, d^3 x .
\ee
Then, obviously, when $\w{C}$ is a conformal Killing vector, the right-hand
side of the above equation vanishes. 
So there exists a solution to Eq.~(\ref{e:ini:DeltaLX_divA}) and
this solution is unique up to the addition of a conformal Killing vector. 
However, given a solution $\w{X}$, for any conformal Killing vector $\w{C}$,
the solution $\w{X}+\w{C}$ yields to the same value of $\w{\tilde L}\w{X}$,
since $\w{C}$ is by definition in the kernel of $\w{\tilde L}$. 
Therefore we conclude that the decomposition (\ref{e:ini:decomp_hA}) of $\hA^{ij}$
is unique, although the vector $\w{X}$ may not be if $(\Sigma_0,\wtgm)$
admits some conformal isometries. 

In the case of an asymptotically flat manifold, the existence and uniqueness
is guaranteed by a theorem proved by Cantor in 1979 \cite{Canto79} 
(see also Appendix~B of Ref.~\cite{SmarrY78a} as well as Refs.~\cite{ChoquIY00,Gourg07a}).
This theorem requires the decay
condition
\be
	\frac{\partial^2\tgm_{ij}}{\partial x^k \partial x^l} = O(r^{-3})	\label{e:ini:aflat_extra}
\ee
in addition to the asymptotic flatness conditions (\ref{e:glob:aflat12}).
This guarantees that
\be
	{\tilde R}_{ij} = O(r^{-3}) . 
\ee
Then all conditions are fulfilled to conclude that 
Eq.~(\ref{e:ini:DeltaLX_divA}) admits a unique solution $\w{X}$ 
which vanishes at infinity. 

To summarize, for all considered cases (asymptotic flatness and
closed manifold), any symmetric and traceless tensor $\hA^{ij}$
(decaying as $O(r^{-2})$ in the asymptotically flat case) 
admits a unique longitudinal/transverse decomposition of
the form (\ref{e:ini:decomp_hA}). 

\subsection{Conformal transverse-traceless form of the constraints}

Inserting the longitudinal/transverse decomposition (\ref{e:ini:decomp_hA})
into the constraint equations (\ref{e:ini:Ham_conf}) and (\ref{e:ini:mom_conf})
and making use of Eq.~(\ref{e:ini:DeltaLX_divA}) yields to the system
\bea
	& & 	 
	\tD_i \tD^i \Psi -\frac{1}{8} {\tilde R} \Psi
	+ \frac{1}{8} \left[(\tilde L X)_{ij} +  \hA_{ij}^{\rm TT}\right]
	 \left[(\tilde L X)^{ij} +  \hA^{ij}_{\rm TT}\right]\, \Psi^{-7}
	\nonumber \\
	& & +  2\pi {\tilde E} \Psi^{-3} - \frac{1}{12} K^2 \Psi^5 = 0  , 
					\label{e:ini:Ham_CTT} \\
	& & 
	 \tilde\Delta_L \, X^i - \frac{2}{3} \Psi^6 \tD^i K 
	= 8\pi {\tilde p}^i  ,	\label{e:ini:mom_CTT}
\eea
where
\be
 (\tilde L X)_{ij} := \tgm_{ik} \tgm_{jl} (\tilde L X)^{kl} 
	\qquad \mbox{and} \qquad
	\hA_{ij}^{\rm TT} := \tgm_{ik} \tgm_{jl} \hA^{kl}_{\rm TT} .
\ee

With the constraint equations written as (\ref{e:ini:Ham_CTT}) and (\ref{e:ini:mom_CTT}),
we see clearly which part of the initial data on $\Sigma_0$ can be freely chosen
and which part is ``constrained'': 
\begin{itemize}
\item free data: 
\begin{itemize}
\item conformal metric $\wtgm$;
\item symmetric traceless and transverse tensor $\hA^{ij}_{\rm TT}$ (traceless and transverse are meant with respect to $\wtgm$: $\tgm_{ij} \hA^{ij}_{\rm TT} = 0$
and $\tD_j \hA^{ij}_{\rm TT} = 0$);
\item scalar field $K$; 
\item conformal matter variables: $(\tilde E,{\tilde p}^i)$;
\end{itemize}
\item constrained data (or ``determined data''): 
\begin{itemize}
\item conformal factor $\Psi$, obeying the non-linear elliptic equation
(\ref{e:ini:Ham_CTT}) (Lichnerowicz equation)
\item vector $\w{X}$, obeying the linear elliptic equation
(\ref{e:ini:mom_CTT}) .
\end{itemize}
\end{itemize}
Accordingly the general strategy to get valid initial data for the Cauchy problem 
is to choose $(\tgm_{ij},\hA^{ij}_{\rm TT},K,\tilde E,{\tilde p}^i)$ on $\Sigma_0$
and solve the system (\ref{e:ini:Ham_CTT})-(\ref{e:ini:mom_CTT}) to get
$\Psi$ and $X^i$. Then one constructs
\bea
	\gm_{ij} & = & \Psi^4 \tgm_{ij} \label{e:ini:recons_gm}\\
	K^{ij} & = & \Psi^{-10} \left( (\tilde L X)^{ij} + \hA^{ij}_{\rm TT} \right)
	+ \frac{1}{3} \Psi^{-4} K \tgm^{ij}  \label{e:ini:recons_K} \\
	E & = & \Psi^{-8} {\tilde E} \\
	p^i & = & \Psi^{-10} {\tilde p}^i 
\eea
and obtains a set $(\wgm,\w{K},E,\w{p})$ which satisfies the constraint
equations (\ref{e:ini:Ham_constr})-(\ref{e:ini:mom_constr}).
This method has been proposed by York (1979) \cite{York79} and is naturally
called the \emph{conformal transverse traceless} (\emph{CTT}) method. 

\subsection{Decoupling on hypersurfaces of constant mean curvature}

Equations (\ref{e:ini:Ham_CTT}) and (\ref{e:ini:mom_CTT}) are coupled, but we
notice that if, among the free data, we choose $K$ to be a constant field
on $\Sigma_0$,
\be \label{e:ini:CMC}
	K = {\rm const},
\ee 
then they decouple partially : condition (\ref{e:ini:CMC}) implies $\tD^i K = 0$,
so that the momentum constraint
(\ref{e:ini:mom_CTT}) becomes independent 
of $\Psi$:
\be \label{e:ini:mom_CMC}
 \tilde\Delta_L \, X^i  = 8\pi {\tilde p}^i \qquad (K={\rm const}) . 
\ee
The condition (\ref{e:ini:CMC}) on the extrinsic curvature of $\Sigma_0$
defines what is called a \emph{constant mean curvature} (\emph{CMC}) hypersurface.
Indeed let us recall that $K$ is nothing but (minus three times) the
mean curvature of $(\Sigma_0,\wgm)$ embedded in $(\mathcal{M},\w{g})$. 
A maximal hypersurface, having $K=0$, is of course a special case of a CMC hypersurface.
On a CMC hypersurface, the task of obtaining initial data is greatly simplified: 
one has first 
to solve the linear elliptic equation (\ref{e:ini:mom_CMC}) to get $\w{X}$
and plug the solution into Eq.~(\ref{e:ini:Ham_CTT}) to form an equation for $\Psi$.
Equation~(\ref{e:ini:mom_CMC}) is the conformal vector Poisson equation 
discussed above (Eq.~(\ref{e:ini:DeltaLX_divA}), with $ \tD_j \hA^{ij} $ replaced
by $8\pi {\tilde p}^i$). We know then that it is 
solvable for the two cases of interest mentioned in 
Sec.~\ref{s:ini:long_trans}: closed or asymptotically flat manifold.
Moreover, the solutions $\w{X}$ are such that the value of $\w{\tilde L} \w{X}$
is unique.

\subsection{Lichnerowicz equation} \label{s:ini:Lichne}

Taking into account the CMC decoupling, 
the difficult problem is to solve Eq.~(\ref{e:ini:Ham_CTT}) for $\Psi$.
This equation is elliptic and highly non-linear\footnote{although it is 
\emph{quasi-linear} in the technical sense, i.e. linear with respect 
to the highest-order derivatives}. 
It has been first studied by Lichnerowicz \cite{Lichn44,Lichn52}
in the case $K=0$ ($\Sigma_0$ maximal) and $\tilde E=0$ (vacuum). 
Lichnerowicz has shown that given 
the value of $\Psi$ at the boundary of a bounded domain of $\Sigma_0$
(Dirichlet problem), there exists at most one solution to Eq.~(\ref{e:ini:Ham_CTT}).
Besides, he showed the existence of a solution provided that $\hA_{ij} \hA^{ij}$
is not too large. 
These early results have been much improved since then. 
In particular  Cantor \cite{Canto77} 
has shown that in the asymptotically flat case, still with
$K=0$ and $\tilde E=0$, Eq.~(\ref{e:ini:Ham_CTT}) is solvable if and only 
if the metric $\wtgm$ is conformal to a metric with vanishing scalar curvature
(one says then that $\wtgm$ belongs to the \emph{positive Yamabe class})
(see also Ref.~\cite{Maxwe04b}).
In the case of closed manifolds, the complete analysis of the CMC case
has been achieved by Isenberg (1995) \cite{Isenb95}.

For more details and further references, we recommend the review articles
by Choquet-Bruhat and York \cite{ChoquY80} and Bartnik and Isenberg \cite{BartnI04}. 
Here we shall simply repeat the argument of York \cite{York99} to 
justify the rescaling (\ref{e:ini:tE_def}) of $E$. This rescaling is indeed 
related to the uniqueness of solutions to the Lichnerowicz equation.
Consider a solution $\Psi_0$ to Eq.~(\ref{e:ini:Ham_CTT}) in the
case $K=0$, to which we restrict ourselves. Another solution close to 
$\Psi_0$ can be written $\Psi=\Psi_0 + \epsilon$, with $|\epsilon| \ll \Psi_0$:
\be
\fl \tD_i \tD^i (\Psi_0+\epsilon) -\frac{{\tilde R}}{8}  (\Psi_0+\epsilon)
	+ \frac{1}{8} \hA_{ij} \hA^{ij} \, (\Psi_0+\epsilon)^{-7}
	+ 2\pi {\tilde E} (\Psi_0+\epsilon)^{-3}  = 0 . 
\ee
Expanding to the first order in $\epsilon/\Psi_0$ leads to 
the following linear equation
for $\epsilon$:
\be \label{e:ini:eq_epsilon}
	\tD_i \tD^i \epsilon - \alpha \epsilon = 0 , 
\ee
with
\be \label{e:ini:def_alpha_Lich}
	\alpha := \frac{1}{8} {\tilde R} 
	+ \frac{7}{8} \hA_{ij} \hA^{ij} \Psi_0^{-8} + 
	6\pi {\tilde E} \Psi_0^{-4} .
\ee
Now, if $\alpha\geq 0$,  one can show, by means of the maximum principle,
that the solution of (\ref{e:ini:eq_epsilon}) which vanishes at spatial
infinity is necessarily $\epsilon=0$ (see Ref.~\cite{ChoquC81} or \S~B.1 of Ref.~\cite{ChoquIY00}). We therefore conclude that the solution $\Psi_0$
to Eq.~(\ref{e:ini:Ham_CTT}) is unique (at least locally) in this case.
On the contrary, if $\alpha<0$, non trivial oscillatory solutions of 
Eq.~(\ref{e:ini:eq_epsilon}) exist, making the solution $\Psi_0$ not unique.
The key point is that the scaling (\ref{e:ini:tE_def}) of $E$ yields the 
term $+6\pi{\tilde E} \Psi_0^{-4}$ in Eq.~(\ref{e:ini:def_alpha_Lich}),
which contributes to make $\alpha$ positive. If we had not rescaled $E$, 
i.e. had considered the original Hamiltonian constraint, the contribution to $\alpha$ would
have been instead $-10\pi E \Psi_0^4$, i.e. would have been negative. 
Actually, any rescaling $\tilde E = \Psi^s E$ with $s>5$ would have work
to make $\alpha$ positive. The choice $s=8$ in Eq.~(\ref{e:ini:tE_def}) 
is motivated by the fact that if the conformal data 
$(\tilde E,\tilde p^i)$ obey the ``conformal''
dominant energy condition
\be
	\tilde E \geq \sqrt{ \tgm_{ij} \tilde p^i \tilde p^j} , 
\ee
then, via the scaling (\ref{e:ini:tE_def}) of $p^i$, the reconstructed physical data
$(E,p^i)$ will automatically obey the dominant energy condition
\be
	E \geq \sqrt{ \gm_{ij} p^i p^j} .
\ee

\section{Conformally flat initial data by the CTT method}

\subsection{Momentarily static initial data}
\label{s:ini:cflat_static}

In this section we search for asymptotically flat initial data 
$(\Sigma_0,\wgm,\w{K})$ by the CTT method exposed above. 
As a purpose of illustration, we shall start by the 
simplest case one may think of, namely choose the freely
specifiable data $(\tgm_{ij},\hA^{ij}_{\rm TT},K,\tilde E,{\tilde p}^i)$ to
be a flat metric:
\be \label{e:ini:tgm_f}
	\tgm_{ij} = f_{ij} ,
\ee
a vanishing transverse-traceless part of the extrinsic curvature:
\be
	\hA^{ij}_{\rm TT} = 0 ,
\ee
a vanishing mean curvature (maximal hypersurface)
\be
	K = 0 ,
\ee 
and a vacuum spacetime:
\be
	\tilde E = 0, \qquad {\tilde p}^i = 0 .
\ee
Then $\tD_i = \Df_i$, where $\wDf$ denotes the Levi-Civita connection associated with
$\w{f}$, $\tilde R = 0$ ($\w{f}$ is flat) and the constraint equations (\ref{e:ini:Ham_CTT})-(\ref{e:ini:mom_CTT})
reduce to 
\bea
	& & \Delta \Psi 
	+ \frac{1}{8} (L X)_{ij} (L X)^{ij} \,  \Psi^{-7} = 0 \label{e:ini:Ham_ex1} \\
	& & \Delta_L X^i = 0 , \label{e:ini:mom_ex1}
\eea
where $\Delta$ and $\Delta_L$ are respectively the scalar Laplacian 
and the conformal vector Laplacian associated with the flat metric $\w{f}$:
\be \label{e:ini:def_Delta}
	\Delta := \Df_i \Df^i
	\qquad \mbox{and} \qquad
	\Delta_L X^i := \Df_j \Df^j X^i + \frac{1}{3} \Df^i \Df_j X^j .
\ee
Equations (\ref{e:ini:Ham_ex1})-(\ref{e:ini:mom_ex1}) must be solved with the boundary conditions
\bea
	& & \Psi = 1 \qquad \mbox{when} \quad r\rightarrow \infty  \label{e:ini:BC_Psi1}\\
	& & \w{X} = 0 \qquad \mbox{when} \quad r\rightarrow \infty , \label{e:ini:BC_X1}
\eea
which follow from the asymptotic flatness requirement. 
The solution depends on the topology of $\Sigma_0$, since the latter may introduce some
inner boundary conditions in addition to (\ref{e:ini:BC_Psi1})-(\ref{e:ini:BC_X1}).

Let us start with the simplest case: $\Sigma_0 = \R^3$. 
Then the unique solution of Eq.~(\ref{e:ini:mom_ex1}) subject to the boundary condition
(\ref{e:ini:BC_X1}) is
\be
	\w{X} = 0 .
\ee
Consequently $(L X)^{ij} = 0$, so that Eq.~(\ref{e:ini:Ham_ex1}) reduces
to Laplace equation for $\Psi$:
\be
	\Delta \Psi = 0 . 
\ee
With the boundary condition (\ref{e:ini:BC_Psi1}), there is a unique regular solution
on $\R^3$: 
\be
	\Psi = 1 . 
\ee
The initial data reconstructed from Eqs.~(\ref{e:ini:recons_gm})-(\ref{e:ini:recons_K})
is then
\bea
	& & \wgm = \w{f}  \\
	& & \w{K} = 0 .
\eea
These data correspond to a spacelike hyperplane of Minkowski spacetime. 
Geometrically the condition $\w{K}=0$ is that of a \emph{totally geodesic hypersurface}
[i.e. all the geodesics of $(\Sigma_t,\wgm)$ are geodesics of $(\mathcal{M},\w{g})$]. Physically data with $\w{K}=0$ 
are said to be \emph{momentarily static} or \emph{time symmetric}. 
Indeed, if we consider a foliation with unit lapse around $\Sigma_0$
(geodesic slicing), the following relation holds: $\Lie{n} \w{g} = - 2 \w{K}$,
where $\Lie{n}$ denotes the Lie derivative along the unit normal $\w{n}$. 
So if $\w{K}=0$, $\Lie{n} \w{g} = 0$. 
This means that, locally (i.e. on $\Sigma_0$), $\w{n}$ is  a spacetime Killing vector.
This vector being timelike, the configuration is then \emph{stationary}.
Moreover, the Killing vector $\w{n}$ being orthogonal to some hypersurface 
(i.e. $\Sigma_0$), the stationary configuration is called \emph{static}.
Of course, this staticity properties holds a priori only on $\Sigma_0$ since 
there is no guarantee that the time development of Cauchy data with $\w{K}=0$
at $t=0$ maintains $\w{K}=0$ at $t>0$. Hence the qualifier \emph{`momentarily'}
in the expression \emph{`momentarily static'} for data with $\w{K}=0$.
\begin{figure}
\centerline{\includegraphics[width=0.8\textwidth]{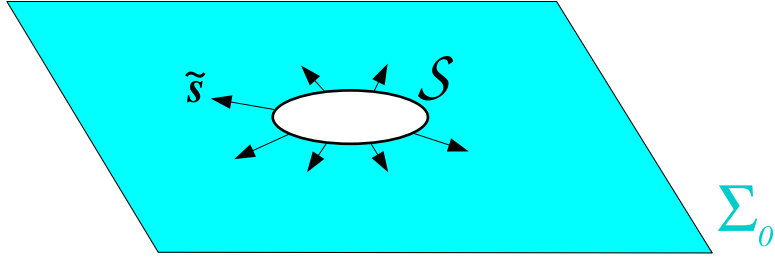}}
\caption[]{\label{f:ini:r3ball} \footnotesize
Hypersurface $\Sigma_0$ as $\R^3$ minus a ball, displayed via an embedding
diagram based on 
the metric $\wtgm$, which coincides with the Euclidean metric on $\R^3$. 
Hence $\Sigma_0$ appears to be flat. 
The unit normal of the inner boundary $\Sp$
with respect to the metric $\wtgm$ 
is $\w{\tilde s}$. Notice that $\wtD\cdot \w{\tilde s}>0$.}
\end{figure}

\subsection{Slice of Schwarzschild spacetime}

\begin{figure}
\centerline{\includegraphics[width=0.8\textwidth]{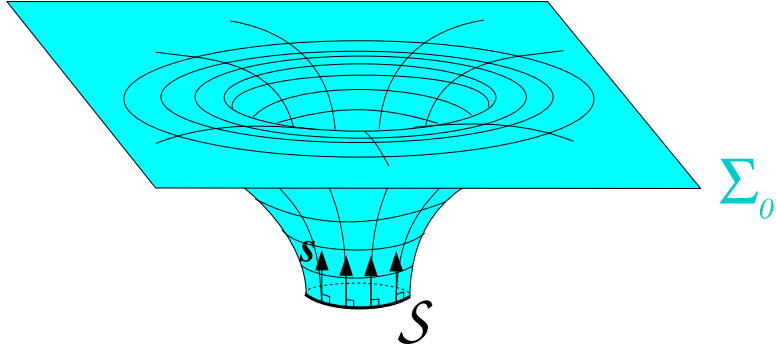}}
\caption[]{\label{f:ini:sminimal} \footnotesize
Same hypersurface $\Sigma_0$ as in Fig.~\ref{f:ini:r3ball} but
displayed via an embedding
diagram based on the metric $\wgm$ instead of $\wtgm$. The unit normal of the inner boundary $\Sp$
with respect to that metric 
is $\w{s}$. Notice that $\wD\cdot \w{s}=0$, which means that $\Sp$ is a 
minimal surface of $(\Sigma_0,\wgm)$.}
\end{figure}

To get something less trivial than a slice of Minkowski spacetime, 
let us consider a slightly more complicated
topology for $\Sigma_0$, namely $\R^3$ minus a ball (cf. Fig.~\ref{f:ini:r3ball}). 
The sphere $\Sp$ delimiting the ball is then the inner boundary of $\Sigma_0$
and we must provide boundary conditions for $\Psi$ and $\w{X}$ on $\Sp$
to solve Eqs.~(\ref{e:ini:Ham_ex1})-(\ref{e:ini:mom_ex1}).
For simplicity, let us choose
\be
	\left.\w{X}\right| _{\Sp} =  0 .
\ee
Altogether with the outer boundary condition (\ref{e:ini:BC_X1}),
this leads to $\w{X}$ being identically zero as the unique solution of Eq.~(\ref{e:ini:mom_ex1}). So, again, the Hamiltonian constraint reduces
to Laplace equation
\be \label{e:ini:Laplace_Psi}
	\Delta \Psi = 0 . 
\ee
If we choose the boundary condition $\left.\Psi\right| _{\Sp} = 1$, then 
the unique solution is $\Psi=1$ and we are back to the previous example
(slice of Minkowski spacetime). 
In order to have something non trivial, i.e. to ensure 
that the metric $\wgm$ will not be flat, let us demand that $\wgm$
admits a \emph{closed minimal surface}, that we will choose to be $\Sp$. 
This will necessarily translate as a boundary condition for $\Psi$ since
all the information on the metric is encoded in $\Psi$ (let us recall that
from the choice (\ref{e:ini:tgm_f}), $\wgm=\Psi^4\w{f}$).
$\Sp$ is a \emph{minimal surface} of $(\Sigma_0,\wgm)$
iff its mean curvature vanishes, or equivalently if its unit normal $\w{s}$
is divergence-free (cf. Fig.~\ref{f:ini:sminimal}):
\be \label{e:ini:div_s_0}
	\left. D_i s^i \right| _{\Sp}= 0 .
\ee
This is the analog of $\w{\nabla}\cdot\w{n}=0$ for maximal hypersurfaces,
the change from \emph{minimal} to \emph{maximal} being due to the change
of metric signature, from the Riemannian to the Lorentzian one. 
Expressed in term of the connection $\wtD = \wDf$ (recall that in the present
case $\wtgm=\w{f}$), condition (\ref{e:ini:div_s_0}) is equivalent to
\be \label{e:ini:div_s_0_1}
	\left. \Df_i (\Psi^6 s^i) \right| _{\Sp}= 0 .
\ee
Let us rewrite this expression in terms of
the unit vector $\w{\tilde s}$ normal to $\Sp$ with respect to the metric $\wtgm$
(cf. Fig.~\ref{f:ini:r3ball}); we have
\be
	\w{\tilde s} = \Psi^{-2} \w{s} ,
\ee
since $\wtgm(\w{\tilde s},\w{\tilde s}) = \Psi^{-4} \wtgm(\w{s},\w{s})
= \wgm(\w{s},\w{s}) = 1$. Thus Eq.~(\ref{e:ini:div_s_0_1}) becomes
\be \label{e:ini:div_s_0_2}
	\left. \Df_i (\Psi^4 {\tilde s}^i) \right| _{\Sp}
 = \left. \frac{1}{\sqrt{f}} \der{}{x^i} \left( \sqrt{f} \Psi^4 {\tilde s}^i \right)
	\right| _{\Sp} = 0 .
\ee
Let us introduce on $\Sigma_0$ a coordinate system of spherical type,
$(x^i)=(r,\theta,\varphi)$, such that (i) $f_{ij} = \mathrm{diag}(1,r^2,r^2\sin^2\theta)$
and (ii) $\Sp$ is the sphere $r=a$, where
$a$ is some positive constant. Since in these coordinates $\sqrt{f} = r^2\sin\theta$ 
and ${\tilde s}^i=(1,0,0)$, the minimal surface condition (\ref{e:ini:div_s_0_2})
is written as
\be
	\left. \frac{1}{r^2} \der{}{r} \left( \Psi^4 r^2 \right) 
		\right| _{r=a}= 0  , 
\ee
i.e. 
\be \label{e:ini:bc_Psi_S}
	\left. \left( \der{\Psi}{r} + \frac{\Psi}{2r} \right) \right| _{r=a} = 0 
\ee
This is a boundary condition of mixed Newmann/Dirichlet type for $\Psi$. 
The unique solution of the Laplace equation (\ref{e:ini:Laplace_Psi}) 
which satisfies boundary conditions
(\ref{e:ini:BC_Psi1}) and (\ref{e:ini:bc_Psi_S}) is
\be
	 \Psi = 1 + \frac{a}{r} . 
\ee
The parameter $a$ is then easily related to the ADM 
mass $m$ of the hypersurface $\Sigma_0$. Indeed for a conformally flat 3-metric
(and more generally in the quasi-isotropic gauge, cf. Chap.~7 of Ref.~\cite{Gourg07a}),
the ADM mass $m$ is given by the flux of the gradient of the conformal factor at spatial
infinity:
\bea 
	m &= &-\frac{1}{2\pi} \lim_{r \rightarrow\infty}
	\oint_{r={\rm const}} 
	 \der{\Psi}{r} r^2 \sin\theta \, d\theta \, d\varphi \nonumber \\
	& = & -\frac{1}{2\pi} \lim_{r \rightarrow\infty}
	 	4\pi r^2 \der{}{r} \left( 1 + \frac{a}{r} \right) 
	= 2 a. \label{e:ini:m_2a}
\eea
Hence $a=m/2$ and we may write
\be
	 \Psi = 1 + \frac{m}{2r} . 
\ee
Therefore, in terms of the 
coordinates $(r,\theta,\varphi)$, the obtained initial data $(\wgm,\w{K})$ are 
\bea
	& & \gm_{ij} = \left( 1 + \frac{m}{2r} \right) ^4
		\mathrm{diag} (1,r^2,r^2\sin\theta) \label{e:ini:gm_Schwarz_iso} \\
	& & K_{ij} = 0 . \label{e:ini:Kij_Schwarz}
\eea
So, as above, the initial data are momentarily static.
Actually, we recognize on (\ref{e:ini:gm_Schwarz_iso})-(\ref{e:ini:Kij_Schwarz})
a slice $t={\rm const}$ of \emph{Schwarzschild spacetime}
in isotropic coordinates.

\begin{figure}
\centerline{\includegraphics[width=0.6\textwidth]{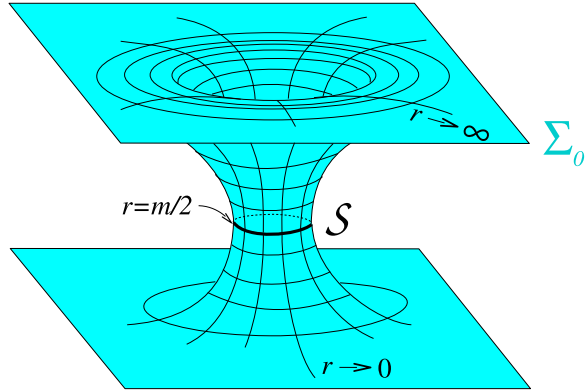}}
\caption[]{\label{f:ini:einst_rosen} \footnotesize
Extended hypersurface $\Sigma'_0$ obtained by gluing a copy of $\Sigma_0$
at the minimal surface $\Sp$; it defines an Einstein-Rosen bridge between
two asymptotically flat regions.}
\end{figure}

\begin{figure}
\centerline{\includegraphics[width=0.6\textwidth]{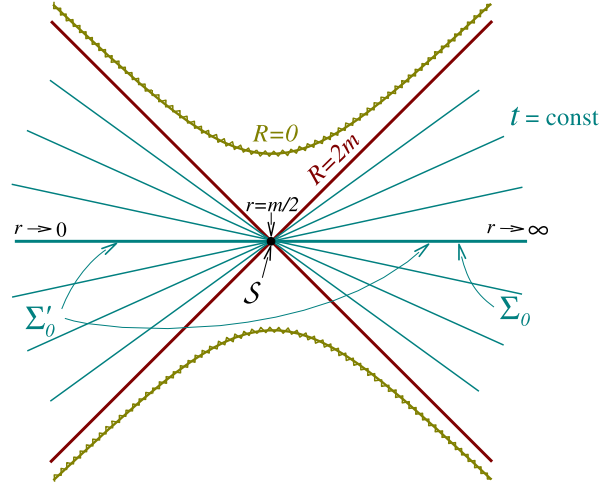}}
\caption[]{\label{f:ini:kruskal} \footnotesize
Extended hypersurface $\Sigma'_0$ depicted in the
Kruskal-Szekeres representation of Schwarzschild spacetime. 
$R$ stands for Schwarzschild radial coordinate and $r$ for the isotropic radial coordinate.
$R=0$ is the singularity and $R=2m$ the event horizon.
$\Sigma'_0$ is nothing but a hypersurface $t={\rm const}$, where $t$ is the
Schwarzschild time coordinate. 
In this diagram, these hypersurfaces are straight lines and the Einstein-Rosen bridge $\Sp$ is reduced to a point.}
\end{figure}

The isotropic coordinates $(r,\theta,\varphi)$ covering the manifold $\Sigma_0$
are such that the range of $r$ is $[m/2,+\infty)$. But thanks to the 
minimal character of the inner boundary $\Sp$, we can extend $(\Sigma_0,\wgm)$ to a
larger Riemannian manifold $(\Sigma'_0,\wgm')$ with
$\left.\wgm'\right| _{\Sigma_0} = \wgm$ and $\wgm'$ smooth at $\Sp$.
This is made possible by gluing a copy of $\Sigma_0$ at $\Sp$
(cf. Fig.~\ref{f:ini:einst_rosen}).
The topology of $\Sigma'_0$ is $\mathbb{S}^2\times \R$ and the
range of $r$ in $\Sigma'_0$ is $(0,+\infty)$. 
The extended metric $\wgm'$ keeps exactly the same form
as (\ref{e:ini:gm_Schwarz_iso}):
\be
	\gm'_{ij}\, dx^i\, dx^j  = \left( 1 + \frac{m}{2r} \right) ^4
	\left( dr^2 + r^2 d\theta^2 + r^2\sin^2\theta d\varphi^2 \right) .
\ee
By the change of variable
\be \label{e:ini:inversion}
	r \mapsto r' = \frac{m^2}{4r}
\ee
it is easily shown that the region $r\rightarrow 0$ does not correspond to some
``center'' but is actually a second asymptotically flat region (the lower one in Fig.~\ref{f:ini:einst_rosen}). Moreover the transformation (\ref{e:ini:inversion}),
with $\theta$ and $\varphi$ kept fixed, is an isometry of $\wgm'$. It maps a
point $p$ of $\Sigma_0$ to the point located at the vertical of $p$ in 
Fig.~\ref{f:ini:einst_rosen}. The minimal sphere $\Sp$ is invariant under
this isometry. The region around $\Sp$
is called an \emph{Einstein-Rosen bridge}. 
$(\Sigma'_0,\wgm')$ is still a slice of Schwarzschild spacetime. 
It connects two asymptotically flat regions without
entering below the event horizon, as shown in the Kruskal-Szekeres diagram 
of Fig.~\ref{f:ini:kruskal}.

\subsection{Bowen-York initial data} \label{s:ini:Bowen_York}

Let us select the same simple free data as above, namely
\be \label{e:ini:BY:free_data}
	\tgm_{ij} = f_{ij}, \quad \hA^{ij}_{\rm TT} = 0, \quad K=0,
	\quad \tilde E = 0
	\quad \mbox{and}\quad {\tilde p}^i = 0 .
\ee
For the hypersurface $\Sigma_0$, instead of $\R^3$ minus a ball, 
we choose $\R^3$ minus a point:
\be
\Sigma_0 = \mathbb{R}^3\backslash\{O\}	.
\ee
The removed point $O$ is called a \emph{puncture} \cite{BrandB97}.
The topology of $\Sigma_0$ is $\mathbb{S}^2\times \R$;
it differs from the topology considered in Sec.~\ref{s:ini:cflat_static}
($\R^3$ minus a ball); actually it is the
same topology as that of the extended manifold $\Sigma'_0$
(cf. Fig.~\ref{f:ini:einst_rosen}).

Thanks to the choice (\ref{e:ini:BY:free_data}), the system to be solved is still (\ref{e:ini:Ham_ex1})-(\ref{e:ini:mom_ex1}). If we choose the trivial solution
$\w{X}=0$ for Eq.~(\ref{e:ini:mom_ex1}), we are back
to the slice of Schwarzschild spacetime considered in Sec.~\ref{s:ini:cflat_static},
except that now $\Sigma_0$ is the extended manifold previously denoted
$\Sigma'_0$. 

Bowen and York \cite{BowenY80} have obtained a simple non-trivial solution 
to the momentum constraint (\ref{e:ini:mom_ex1}) (see also Ref.~\cite{BeigK04}).
Given a Cartesian coordinate system $(x^i)=(x,y,z)$ on $\Sigma_0$ 
(i.e. a coordinate system such that $f_{ij}=\mathrm{diag}(1,1,1)$) with respect
to which the coordinates of the puncture $O$ are $(0,0,0)$, this solution 
writes 
\be \label{e:ini:BY_X}
X^i = - \frac{1}{4 r} \left( 7 f^{ij} P_j +  \frac{P_j x^j x^i }{r^2}
	\right) - \frac{1}{r^3} \epsilon^{ij}_{\ \ k} S_j x^k ,
\ee
where $r:=\sqrt{x^2+y^2+z^2}$, $\epsilon^{ij}_{\ \ k}$ is the Levi-Civita alternating
tensor associated with the flat metric $\w{f}$ and 
$(P_i,S_j)=(P_1,P_2,P_3,S_1,S_2,S_3)$
are six real numbers, which constitute the six parameters of the Bowen-York solution.
Notice that since $r\not=0$ on $\Sigma_0$, the Bowen-York solution is a regular 
and smooth solution on the entire $\Sigma_0$.

The conformal traceless extrinsic curvature corresponding to the solution
(\ref{e:ini:BY_X}) is deduced from formula (\ref{e:ini:decomp_hA}), which in 
the present case reduces to $\hA^{ij} = (L X)^{ij}$; one gets
\be
 \fl \hA^{ij} = \frac{3}{2 r^3} \left[ x^i P^j + x^j P^i  - \left(
			f^{ij} - \frac{x^i x^j}{r^2} \right)
			P_k x^k \right]
		+ \frac{3}{r^5} \left( \epsilon^{ik}_{\ \  l} S_k x^l x^j
			+ \epsilon^{jk}_{\ \  l} S_k x^l x^i \right) ,\label{e:ini:BY_hA}
\ee
where $P^i := f^{ij} P_j$.
The tensor $\hA^{ij}$ given by Eq.~(\ref{e:ini:BY_hA}) is called the 
\emph{Bowen-York extrinsic curvature}. Notice that 
the $P_i$ part of $\hA^{ij}$ decays asymptotically as $O(r^{-2})$, whereas the
$S_i$ part decays as $O(r^{-3})$.
\begin{remark}
Actually the expression of $\hA^{ij}$ given in the original Bowen-York article 
\cite{BowenY80} contains an additional term with respect to Eq.~(\ref{e:ini:BY_hA}),
but the role of this extra term is only to ensure that the solution is isometric
through an inversion across some sphere. We are not interested by such a property
here, so we have dropped this term. Therefore, strictly speaking, we should
name expression (\ref{e:ini:BY_hA}) the \emph{simplified} Bowen-York extrinsic curvature.
\end{remark}

The Bowen-York extrinsic curvature provides an analytical solution of the
momentum constraint (\ref{e:ini:mom_ex1}) but there remains to solve the
Hamiltonian constraint (\ref{e:ini:Ham_ex1}) for $\Psi$, with
the asymptotic flatness boundary condition $\Psi=1$ when $r\rightarrow \infty$. 
Since $\w{X}\not=0$, Eq.~(\ref{e:ini:Ham_ex1}) is no
longer a simple Laplace equation, as in Sec.~\ref{s:ini:cflat_static}, but a
 non-linear elliptic equation. There is no hope to get any analytical
solution and one must solve Eq.~(\ref{e:ini:Ham_ex1}) numerically to get $\Psi$
and reconstruct the full initial data $(\wgm,\w{K})$ via 
Eqs.~(\ref{e:ini:recons_gm})-(\ref{e:ini:recons_K}).

The parameters $P_i$ of the Bowen-York solution are nothing but the
three components of the ADM linear momentum of the hypersurface $\Sigma_0$
Similarly, the parameters $S_i$ of the Bowen-York solution are nothing but the
three components of the angular momentum of the hypersurface $\Sigma_0$,
the latter being defined relatively to the quasi-isotropic gauge, in the
absence of any axial symmetry (see e.g. \cite{Gourg07a}).

\begin{remark}
The Bowen-York solution with $P^i=0$ and $S^i=0$ reduces to the
momentarily static  solution found in Sec.~\ref{s:ini:cflat_static}, i.e. 
is a slice $t={\rm const}$ of the Schwarzschild spacetime ($t$ being the
Schwarzschild time coordinate). 
However Bowen-York initial data with $P^i=0$ and $S^i\not=0$ do not 
constitute a slice of Kerr spacetime. Indeed, it has been
shown \cite{GaratP00} that there does not exist any foliation of Kerr spacetime by hypersurfaces
which (i) are
axisymmetric, (ii) smoothly reduce in the non-rotating limit to the hypersurfaces
of constant Schwarzschild time and (iii) are conformally flat, i.e. 
have induced metric $\wtgm=\w{f}$, as the Bowen-York hypersurfaces have. 
This means that a Bowen-York solution with $S^i\not=0$ does represent 
initial data for a rotating black hole, but this black hole is not stationary:
it is ``surrounded'' by gravitational radiation, as demonstrated by the
time development of these initial data \cite{BrandS95b,GleisNPP98}.
\end{remark}

\section{Conformal thin sandwich method}

\subsection{The original conformal thin sandwich method} \label{s:ini:CTS_ori}

An alternative to the conformal transverse-traceless method for computing
initial data has been introduced by York in 1999 \cite{York99}. 
The starting point is the identity
\be \label{e:K_kin}
	\w{K} = - \frac{1}{2N} \w{\mathcal{L}}_{N\w{n}} \wgm 
	= - \frac{1}{2N} \left( \der{}{t} - \Lie{\beta} \right) \wgm, 
\ee
where $N$ is the lapse function and $\w{\beta}$ is the shift vector associated with
some 3+1 coordinates $(t,x^i)$. The traceless part of Eq.~(\ref{e:K_kin}) leads to 
\be \label{e:ini:tA_Lmg}
	\tA^{ij} = \frac{1}{2N} \left[ \left( \der{}{t} - \Liec{\beta} \right)
	 \tgm^{ij}  - \frac{2}{3} \tD_k \beta^k \, \tgm^{ij}  \right] ,
\ee
where $\tA^{ij}$ is defined by Eq.~(\ref{e:def_tA}). 
Noticing that
\be
- \Liec{\beta} \tgm^{ij} = (\tilde L \beta)^{ij}+ \frac{2}{3} \tD_k \beta^k , 
\ee
and introducing the short-hand notation
\be
	\dot\tgm^{ij} := \der{}{t}\tgm^{ij} , 
\ee
we can rewrite Eq.~(\ref{e:ini:tA_Lmg}) as
\be  \label{e:ini:tA_Lbeta}
	\tA^{ij} = \frac{1}{2N} \left[ \dot\tgm^{ij} + (\tilde L \beta)^{ij} \right] .
\ee
The relation between $\tA^{ij}$ and $\hA^{ij}$ is 
[cf. Eqs.~(\ref{e:def_hA})-(\ref{e:def_tA})]
\be
	\hA^{ij} = \Psi^6 \tA^{ij} . 
\ee
Accordingly, Eq.~(\ref{e:ini:tA_Lbeta}) yields
\be \label{e:ini:hA_dg_beta}
   \hA^{ij} = \frac{1}{2\tilde N} \left[ \dot\tgm^{ij} + 
	(\tilde L \beta)^{ij} \right]  , 
\ee
where we have introduced the \emph{conformal lapse}
\be \label{e:ini:def_tN}
	 \tilde N := \Psi^{-6} N . 
\ee
Equation~(\ref{e:ini:hA_dg_beta}) constitutes a decomposition of $\hA^{ij}$
alternative to the longitudinal/transverse decomposition (\ref{e:ini:decomp_hA}). 
Instead of expressing $\hA^{ij}$ in terms of a vector $\w{X}$ and
a TT tensor $\hA^{ij}_{\rm TT}$, it expresses it in terms of the shift vector
$\w{\beta}$, the time derivative of the conformal metric, $\dot\tgm^{ij}$, 
and the conformal lapse $\tilde N$. 

The Hamiltonian constraint, written as the Lichnerowicz equation
(\ref{e:ini:Ham_conf}), takes the same form as before:
\be \label{e:ini:Ham_CTS}
\tD_i \tD^i \Psi -\frac{{\tilde R}}{8}  \Psi
	+ \frac{1}{8} \hA_{ij} \hA^{ij} \, \Psi^{-7}
	+ 2\pi {\tilde E} \Psi^{-3} - \frac{K^2}{12}  \Psi^5 = 0 ,	
\ee
except that now $\hA^{ij}$ is to be understood as the
combination (\ref{e:ini:hA_dg_beta}) of $\beta^i$, $\dot\tgm^{ij}$ and
$\tilde N$. 
On the other side, the momentum constraint (\ref{e:ini:mom_conf}) becomes, 
once expression (\ref{e:ini:hA_dg_beta}) is substituted for $\hA^{ij}$, 
\be \label{e:ini:mom_CTS}
	 \tD_j \left( \frac{1}{\tilde N} (\tilde L \beta)^{ij} \right)
 + \tD_j \left( \frac{1}{\tilde N} \dot\tgm^{ij} \right)
  - \frac{4}{3} \Psi^6 \tD^i K = 16\pi {\tilde p}^i . 
\ee
In view of the system (\ref{e:ini:Ham_CTS})-(\ref{e:ini:mom_CTS}), the 
method to compute initial data consists in choosing freely 
$\tgm_{ij}$,  $\dot\tgm^{ij}$, $K$, ${\tilde N}$, ${\tilde E}$
and ${\tilde p}^i$ on $\Sigma_0$ and solving (\ref{e:ini:Ham_CTS})-(\ref{e:ini:mom_CTS}) to get $\Psi$ and $\beta^i$. 
This method is called \emph{conformal thin sandwich} (\emph{CTS}), 
because one input
is the time derivative $\dot\tgm^{ij}$, which can be obtained from 
the value of the conformal metric on two neighbouring hypersurfaces
$\Sigma_t$ and $\Sigma_{t+\delta t}$ (``thin sandwich'' view point). 

\begin{remark}
The term ``thin sandwich'' originates from a previous method 
devised in the early sixties by Wheeler and his collaborators 
\cite{BaierSW62,Wheel64}. Contrary to the methods exposed here, the
thin sandwich method
was not based on a conformal decomposition: it considered the constraint
equations (\ref{e:ini:Ham_constr})-(\ref{e:ini:mom_constr}) as a system
to be solved for the lapse $N$ and the shift vector $\w{\beta}$, given
the metric $\wgm$ and its time derivative. The extrinsic curvature 
which appears in (\ref{e:ini:Ham_constr})-(\ref{e:ini:mom_constr}) was
then considered as the function of $\wgm$, $\dert{\wgm}{t}$, $N$ and $\w{\beta}$
given by Eq.~(\ref{e:K_kin}). However, this method does
not work in general \cite{BartnF93}. On the contrary the \emph{conformal} 
thin sandwich method introduced by York \cite{York99} and exposed
above was shown to work
\cite{ChoquIY00}. 
\end{remark}

As for the conformal transverse-traceless method treated in Sec.~\ref{s:ini:CTT}, 
on CMC hypersurfaces, Eq.~(\ref{e:ini:mom_CTS})
decouples from Eq.~(\ref{e:ini:Ham_CTS}) and becomes an elliptic linear equation 
for $\w{\beta}$. 

\subsection{Extended conformal thin sandwich method}

An input of the above method is the conformal lapse $\tilde N$.
Considering the astrophysical problem stated in Sec.~\ref{s:ini:idp},
it is not clear how to pick a relevant value for $\tilde N$.
Instead of choosing an arbitrary value, 
Pfeiffer and York \cite{PfeifY03} have suggested
to compute $\tilde N$ from the Einstein equation
giving the time derivative of the trace $K$ of the extrinsic curvature,
i.e. 
\bea 
 \left(\der{}{t} - \Liec{\beta} \right) K 
	& = & - \Psi^{-4} \left( \tD_i \tD^i N + 2 \tD_i \ln \Psi \, \tD^i N \right) 
	 \nonumber \\
	 & & + N \left[ 4\pi (E+S) 
	+  \tA_{ij} \tA^{ij} + \frac{K^2}{3}\right] , \label{e:ini:evol_K}
\eea
where $S$ is the trace of the matter stress tensor as measured by the Eulerian
observer: $S = \gm^{\mu\nu} T_{\mu\nu}$. 
This amounts to add this equation to the initial data system. 
More precisely, Pfeiffer and York \cite{PfeifY03} suggested to combine 
Eq.~(\ref{e:ini:evol_K}) with the Hamiltonian constraint to get an equation
involving the quantity $N\Psi = \tilde N \Psi^7$ and containing no
scalar products of gradients as the $\tD_i \ln\Psi \tD^i N$ term in 
Eq.~(\ref{e:ini:evol_K}), thanks to the identity
\be
	\tD_i \tD^i N + 2 \tD_i \ln \Psi \, \tD^i N 
	= \Psi^{-1} \left[ \tD_i \tD^i (N\Psi) + N \tD_i \tD^i \Psi \right] . 
\ee
Expressing the left-hand side of the above equation in terms of Eq.~(\ref{e:ini:evol_K})
and substituting $\tD_i \tD^i \Psi$ in the right-hand side by its expression
deduced from Eq.~(\ref{e:ini:Ham_CTS}), we get 
\bea 
 \fl \tD_i \tD^i (\tilde N \Psi^7) - (\tilde N \Psi^7)
	\left[ \frac{1}{8} \tilde R + \frac{5}{12} K^2 \Psi^4
	+ \frac{7}{8} \hA_{ij} \hA^{ij} \Psi^{-8} 
	+ 2 \pi (\tilde E+2\tilde S) \Psi^{-4} \right] \nonumber \\
	+ \left( \dot K - \beta^i \tD_i K \right) \Psi^5 = 0 ,\label{e:ini:eq_NPsi7}
\eea
where we have used the short-hand notation
\be
	\dot K := \der{K}{t} 
\ee
and have set
\be
	\tilde S := \Psi^8 S . 
\ee
Adding Eq.~(\ref{e:ini:eq_NPsi7}) to Eqs.~(\ref{e:ini:Ham_CTS}) and
(\ref{e:ini:mom_CTS}), the initial data system becomes
\bea
 \fl \tD_i \tD^i \Psi -\frac{{\tilde R}}{8}  \Psi
	+ \frac{1}{8} \hA_{ij} \hA^{ij} \, \Psi^{-7}
	+ 2\pi {\tilde E} \Psi^{-3} - \frac{K^2}{12}  \Psi^5 = 0  \label{e:ini:XCTS1} \\
 \fl  \tD_j \left( \frac{1}{\tilde N} (\tilde L \beta)^{ij} \right)
 + \tD_j \left( \frac{1}{\tilde N} \dot\tgm^{ij} \right)
  - \frac{4}{3} \Psi^6 \tD^i K = 16\pi {\tilde p}^i  \label{e:ini:XCTS2} \\
 \fl  \tD_i \tD^i (\tilde N \Psi^7) - (\tilde N \Psi^7)
	\bigg[ \frac{\tilde R}{8} + \frac{5}{12} K^2 \Psi^4
	+ \frac{7}{8} \hA_{ij} \hA^{ij} \Psi^{-8} 
	+ 2 \pi (\tilde E+2\tilde S) \Psi^{-4} \bigg] \nonumber \\ 
	 + \left( \dot K - \beta^i \tD_i K \right) \Psi^5 = 0   , \label{e:ini:XCTS3}
\eea
where $\hA^{ij}$ is the function of $\tilde N$, $\beta^i$, $\tgm_{ij}$ and
$\dot\tgm^{ij}$ defined by Eq.~(\ref{e:ini:hA_dg_beta}).
Equations (\ref{e:ini:XCTS1})-(\ref{e:ini:XCTS3}) constitute the
\emph{extended conformal thin sandwich} (\emph{XCTS}) 
system for the initial data problem.
The free data are the conformal metric $\wtgm$, its coordinate time derivative
$\w{\dot\tgm}$, the extrinsic curvature trace $K$, its coordinate time derivative
$\dot K$, and the rescaled matter variables $\tilde E$, $\tilde S$ and 
$\tilde p^i$. The constrained data are the conformal factor $\Psi$, the 
conformal lapse $\tilde N$ and the shift vector $\w{\beta}$. 
\begin{remark}
The XCTS system (\ref{e:ini:XCTS1})-(\ref{e:ini:XCTS3}) is a coupled system.
Contrary to the CTT system (\ref{e:ini:Ham_CTT})-(\ref{e:ini:mom_CTT}), the
assumption of constant mean curvature, and in particular of maximal slicing,
does not allow to decouple it.
\end{remark}

\subsection{XCTS at work: static black hole example} \label{s:ini:XCTS_work}

Let us illustrate the extended conformal thin sandwich method on a simple
example. Take for the hypersurface $\Sigma_0$ the punctured manifold considered
in Sec.~\ref{s:ini:Bowen_York}, namely
\be
\Sigma_0 = \R^3\backslash\{O\}	.
\ee
For the free data, let us perform the simplest choice:
\be \label{e:ini:XCTS_ex_free}
	\fl \tgm_{ij} = f_{ij}, 
	\quad \dot\tgm^{ij} = 0, \quad K=0, \quad \dot K =0,
	\quad \tilde E = 0,
	\quad \tilde S = 0,
	\quad \mbox{and}\quad {\tilde p}^i = 0 ,
\ee
i.e. we are searching for vacuum initial data on a maximal and conformally flat 
hypersurface with all the freely specifiable time derivatives set to zero.
Thanks to (\ref{e:ini:XCTS_ex_free}), the XCTS system (\ref{e:ini:XCTS1})-(\ref{e:ini:XCTS3}) reduces to
\bea
 & & \Delta \Psi 
	+ \frac{1}{8} \hA_{ij} \hA^{ij} \, \Psi^{-7} = 0 \label{e:ini:XCTS_ex1} \\
 & & \Df_j \left( \frac{1}{\tilde N} (L \beta)^{ij} \right) = 0 \label{e:ini:XCTS_ex2} \\
& & \Delta (\tilde N \Psi^7) - 
	\frac{7}{8} \hA_{ij} \hA^{ij} \Psi^{-1} \tilde N 
		= 0 \label{e:ini:XCTS_ex3} .
\eea
Aiming at finding the simplest solution, we notice that
\be \label{e:ini:XCTS_ex_beta}
	\w{\beta} = 0
\ee
is a solution of Eq.~(\ref{e:ini:XCTS_ex2}). Together with $\dot\tgm^{ij} = 0$, it 
leads to [cf. Eq.~(\ref{e:ini:hA_dg_beta})]
\be
	\hA^{ij} = 0 .
\ee
The system (\ref{e:ini:XCTS_ex1})-(\ref{e:ini:XCTS_ex3}) reduces then further:
\bea
  	& & \Delta \Psi = 0 \label{e:ini:XCTS_ex1-2} \\
	& & \Delta (\tilde N \Psi^7) = 0 . \label{e:ini:XCTS_ex3-2}
\eea
Hence we have only two Laplace equations to solve. Moreover Eq.~(\ref{e:ini:XCTS_ex1-2})
decouples from Eq.~(\ref{e:ini:XCTS_ex3-2}). For simplicity, let us assume spherical
symmetry around the puncture $O$. 
We introduce an adapted spherical coordinate
system $(x^i) = (r,\theta,\varphi)$ on $\Sigma_0$.
The puncture $O$ is then at $r=0$. The
simplest non-trivial solution of (\ref{e:ini:XCTS_ex1-2}) which obeys the asymptotic
flatness condition $\Psi\rightarrow 1$ as $r\rightarrow+\infty$ is 
\be \label{e:ini:XCTS_ex_Psi}
	\Psi = 1 + \frac{m}{2r} , 
\ee
where as in Sec.~\ref{s:ini:cflat_static}, the constant $m$ is the ADM mass of $\Sigma_0$
[cf. Eq.~(\ref{e:ini:m_2a})]. Notice that since $r=0$ is excluded from $\Sigma_0$,
$\Psi$ is a perfectly regular solution on the entire manifold $\Sigma_0$.
Let us recall that the Riemannian manifold $(\Sigma_0,\wgm)$ corresponding to this
value of $\Psi$ via $\wgm=\Psi^4 \w{f}$ is the Riemannian manifold denoted $(\Sigma'_0,\wgm)$ in Sec.~\ref{s:ini:cflat_static} and depicted in 
Fig.~\ref{f:ini:einst_rosen}. In particular it has two asymptotically flat ends:
$r\rightarrow+\infty$ and $r\rightarrow 0$ (the puncture). 

As for Eq.~(\ref{e:ini:XCTS_ex1-2}), 
the simplest solution of Eq.~(\ref{e:ini:XCTS_ex3-2}) obeying the
asymptotic flatness requirement $\tilde N\Psi^7 \rightarrow 1$ as 
$r\rightarrow+\infty$ is
\be \label{e:ini:ex_tN_Psi7}
	\tilde N \Psi^7 = 1 + \frac{a}{r} , 
\ee
where $a$ is some constant.
Let us determine $a$ from the value of the lapse function at the second
asymptotically flat end $r\rightarrow 0$. The lapse being related to $\tilde N$
via Eq.~(\ref{e:ini:def_tN}), Eq.~(\ref{e:ini:ex_tN_Psi7}) is equivalent
to
\be \label{e:ini:XCTS_ex_N}
	N = \left( 1 + \frac{a}{r} \right) \Psi^{-1} 
	=  \left(1 + \frac{a}{r} \right) \left( 1 + \frac{m}{2r} \right) ^{-1} 
	= \frac{r+a}{r+m/2} . 
\ee
Hence
\be  \label{e:ini:XCTS_ex_limN_a}
	\lim_{r\rightarrow 0} N = \frac{2a}{m} . 
\ee
There are two natural choices for $\lim_{r\rightarrow 0} N$. The first one
is 
\be \label{e:ini:XCTS_ex_limN_1}
  \lim_{r\rightarrow 0} N = 1,
\ee
 yielding $a=m/2$. Then, from Eq.~(\ref{e:ini:XCTS_ex_N})
$N=1$ everywhere on $\Sigma_0$. This value of $N$ corresponds to a geodesic
slicing. The second choice is
\be \label{e:ini:XCTS_ex_limN_m1}
	\lim_{r\rightarrow 0} N = -1. 
\ee
This choice is compatible with asymptotic flatness: it simply means that
the coordinate time $t$ is running ``backward'' near the asymptotic flat end
$r\rightarrow 0$. This contradicts the assumption $N>0$ in 
the standard definition of the lapse function. 
However, we shall generalize here the definition of the lapse to allow for
negative values: whereas the unit vector $\w{n}$ is always future-oriented, the
scalar field $t$ is allowed to decrease towards the future. Such a situation 
has already been encountered for the part of the slices $t={\rm const}$ located on the
left side of Fig.~\ref{f:ini:kruskal}. Once reported 
into Eq.~(\ref{e:ini:XCTS_ex_limN_a}),
the choice (\ref{e:ini:XCTS_ex_limN_m1}) yields $a=-m/2$, so that
\be \label{e:ini:XCTS_ex_N-2}
	N = \left(1 - \frac{m}{2r} \right) \left( 1 + \frac{m}{2r} \right) ^{-1} .
\ee
Gathering relations (\ref{e:ini:XCTS_ex_beta}), (\ref{e:ini:XCTS_ex_Psi}) and
(\ref{e:ini:XCTS_ex_N-2}), we arrive at the following expression of the spacetime metric
components:
\be \label{e:ini:XCTS_ex_gab}
	\fl   g_{\mu\nu} dx^\mu dx^\nu  = - \left( 
    \frac{1 - \frac{m}{2r}}{ 1 + \frac{m}{2r}} \right) ^2
         dt^2 
    + \left( 1 + \frac{m}{2r} \right) ^4 \left[ d{r}^2 
    + {r}^2 (d\theta^2 + \sin^2\theta d\varphi^2) \right]  .
\ee
We recognize the line element of Schwarzschild spacetime in isotropic coordinates. 
Hence we recover the same initial
data as in Sec.~\ref{s:ini:cflat_static} and depicted in Figs.~\ref{f:ini:einst_rosen}
and \ref{f:ini:kruskal}. The bonus is that we have the complete expression
of the metric $\w{g}$ on $\Sigma_0$, and not only the induced metric $\wgm$.
\begin{remark}
The choices (\ref{e:ini:XCTS_ex_limN_1}) and (\ref{e:ini:XCTS_ex_limN_m1})
for the asymptotic value of the lapse both lead to a momentarily static
initial slice in Schwarzschild spacetime. The difference is that the
time development corresponding to choice (\ref{e:ini:XCTS_ex_limN_1}) 
(geodesic slicing) will depend on $t$, whereas the time development corresponding to choice (\ref{e:ini:XCTS_ex_limN_m1}) will not, since in the latter case $t$
coincides with the standard Schwarzschild time coordinate, 
which makes $\wpar_t$ a Killing vector.
\end{remark}

\subsection{Uniqueness of solutions}

Recently, Pfeiffer and York \cite{PfeifY05} have exhibited a choice of vacuum free
data $(\tgm_{ij},\dot\tgm^{ij},K,\dot K)$ for which the solution 
$(\Psi,\tilde N,\beta^i)$ to the XCTS system (\ref{e:ini:XCTS1})-(\ref{e:ini:XCTS3}) 
is not unique (actually two solutions are found). 
The conformal metric $\wtgm$ is the flat metric plus 
a linearized quadrupolar gravitational wave, as obtained by Teukolsky 
\cite{Teuko82}, with a tunable amplitude. $\dot\tgm^{ij}$ corresponds to the
time derivative of this wave, and both $K$ and $\dot K$ are chosen to zero.
On the contrary, for the same free data, with $\dot K=0$ substituted by 
$\tilde N=1$, Pfeiffer and York have shown that the original conformal
thin sandwich method as described in Sec.~\ref{s:ini:CTS_ori} leads to a unique
solution (or no solution at all if the amplitude of the wave is two large). 

Baumgarte, \'O Murchadha and Pfeiffer \cite{BaumgOP06} have argued that the lack
of uniqueness for the XCTS system may be due to the term
\be
	\fl -(\tilde N \Psi^7) \frac{7}{8} \hA_{ij} \hA^{ij} \Psi^{-8} = 
	 - \frac{7}{32} \Psi^6 \tgm_{ik} \tgm_{jl}
	\left[ \dot\tgm^{ij} + (\tilde L \beta)^{ij} \right]
	\left[ \dot\tgm^{kl} + (\tilde L \beta)^{kl} \right]
	\, (\tilde N \Psi^7)^{-1}
\ee
in Eq.~(\ref{e:ini:XCTS3}). Indeed, if we proceed as for the analysis of
Lichnerowicz equation in Sec.~\ref{s:ini:Lichne}, we notice that this
term, with the minus sign and the
negative power of $(\tilde N \Psi^7)^{-1}$, makes the linearization of Eq.~(\ref{e:ini:XCTS3}) of the type $\tD_i \tD^i \epsilon + \alpha\epsilon=\sigma$,
with $\alpha>0$. This ``wrong'' sign of $\alpha$ prevents the application of the
maximum principle to guarantee the uniqueness of the solution. 

The non-uniqueness of solution of the XCTS system for certain choice of 
free data has been confirmed by Walsh \cite{Walsh06} by means of bifurcation
theory.

\subsection{Comparing CTT, CTS and XCTS}

The conformal transverse traceless (CTT) method exposed in Sec.~\ref{s:ini:CTT}
and the (extended) conformal thin sandwich (XCTS) method considered here
differ by the choice of free data:
whereas both methods use the conformal metric
$\wtgm$ and the trace of the extrinsic curvature $K$ as free data,
CTT employs in addition $\hA^{ij}_{\rm TT}$, whereas for CTS (resp. XCTS) 
the additional
free data is $\dot\tgm^{ij}$, as well as $\tilde N$ (resp. $\dot K$).
Since $\hA^{ij}_{\rm TT}$ is directly related to the extrinsic curvature
and the latter is linked to the canonical momentum of the gravitational field
in the Hamiltonian formulation of general relativity, 
the CTT method can be considered as the approach to the initial data problem
in the \emph{Hamiltonian representation}. On the other side, $\dot\tgm^{ij}$ being
the ``velocity'' of $\tgm^{ij}$, the (X)CTS method constitutes the approach
in the \emph{Lagrangian representation} \cite{York04}.

\begin{remark}
The (X)CTS method assumes that the conformal metric
is unimodular: $\det (\tgm_{ij}) = f$
(since Eq.~(\ref{e:ini:hA_dg_beta}) follows from this assumption),
whereas the CTT method can be applied with any
conformal metric. 
\end{remark}

The advantage of CTT is that its mathematical theory is well developed, 
yielding existence and uniqueness theorems, at least for constant mean curvature
(CMC) slices. The mathematical theory of CTS is very close to CTT. In particular, the
momentum constraint decouples from the Hamiltonian constraint on CMC slices. 
On the contrary, XCTS has a much more involved mathematical structure. In
particular the CMC condition does not yield to any decoupling. 
The advantage of XCTS is then to be better suited to the description of
quasi-stationary spacetimes, since 
$\dot\tgm^{ij}=0$ and $\dot K=0$ are necessary conditions for 
$\wpar_t$ to be a  Killing vector. This makes XCTS the method to be used in 
order to prepare initial data in quasi-equilibrium.
For instance, it has been shown \cite{GrandGB02,DamouGG02}
that XCTS yields orbiting binary black hole configurations 
in much better agreement with post-Newtonian computations
than the CTT treatment based on a superposition of two Bowen-York solutions. 
Indeed, except when they are very close and about to merge, the orbits
of binary black holes evolve very slowly, so that it is a very good approximation 
to consider that the system is in quasi-equilibrium. XCTS takes this fully into
account, while CTT relies on a technical simplification (Bowen-York analytical solution
of the momentum constraint), with no direct relation to the quasi-equilibrium state. 

A detailed comparison of CTT and XCTS for a single spinning or boosted 
black hole has been performed by Laguna \cite{Lagun04}.


\section{Initial data for binary systems} \label{s:ini:binary}

A major topic of contemporary numerical relativity is the computation of the 
merger of a binary system of black holes \cite{Campa07} or neutron stars \cite{Shiba07}, 
for such systems
are among the most promising sources of gravitational radiation for the
interferometric detectors either groundbased (LIGO, VIRGO, GEO600, TAMA) or 
in space (LISA). 
The problem of preparing initial data for these systems has therefore  
received a lot of attention in the past decade. 

\begin{figure}
\centerline{\includegraphics[width=0.6\textwidth]{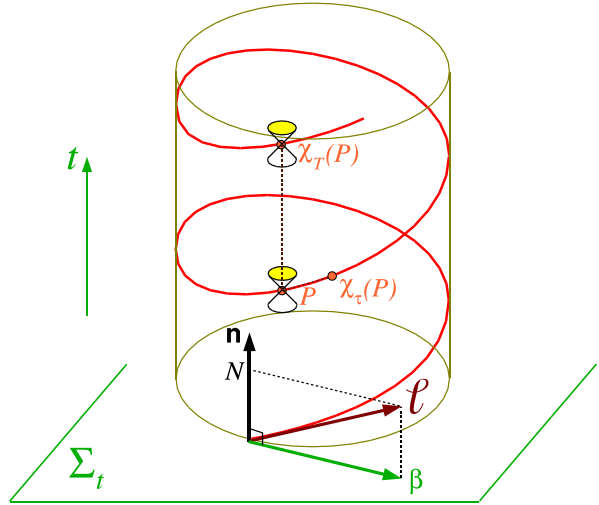}}
\caption[]{\label{f:ini:helical} \footnotesize
Action of the helical symmetry group, with Killing vector $\w{\ell}$.
$\chi_\tau(P)$ is the displacement of the point $P$ 
by the member of the symmetry group of parameter $\tau$.
$N$ and $\w{\beta}$ are respectively the lapse function and the shift vector
associated with coordinates adapted to the symmetry, i.e. coordinates $(t,x^i)$
such that $\wpar_t = \w{\ell}$.}
\end{figure}

\subsection{Helical symmetry}

Due to the gravitational-radiation reaction, a relativistic binary system 
has an inspiral
motion, leading to the merger of the two components. However, when the 
two bodies are sufficiently far apart, one may approximate the spiraling
orbits by closed ones. Moreover, it is well known that gravitational radiation circularizes the orbits very efficiently, at least for comparable mass systems
\cite{Blanc06a}. 
We may then consider that the motion is described by a
sequence of \emph{closed circular orbits}. 

The geometrical translation of this physical assumption is that the
spacetime $(\mathcal{M},\w{g})$ is endowed with some symmetry, 
called \emph{helical symmetry}.
Indeed exactly circular orbits imply the existence of a one-parameter symmetry
group such that the associated Killing vector $\w{\ell}$ obeys 
the following properties \cite{FriedUS02}:
(i) $\w{\ell}$ is timelike near the system, (ii) far from it, $\w{\ell}$ is spacelike
but there exists a smaller number $T>0$ such that the separation between any point 
$P$ and its image $\chi_T(P)$ under the symmetry group is timelike
(cf. Fig.~\ref{f:ini:helical}). $\w{\ell}$ is called a 
\emph{helical Killing vector}, its field lines in a spacetime 
diagram being helices (cf. Fig.~\ref{f:ini:helical}).

Helical symmetry is exact in theories of gravity where gravitational radiation
does not exist, namely: 
\begin{itemize}
\item in Newtonian gravity,
\item in post-Newtonian gravity, up to the second order,
\item in the Isenberg-Wilson-Mathews (IWM) approximation to general relativity, based
on the assumptions $\wtgm = \w{f}$  and $K=0$ \cite{Isenb78,WilsoM89}.
\end{itemize}
Moreover helical symmetry can be exact in full general relativity for a 
non-axisymmetric system (such
as a binary) with standing gravitational waves \cite{Detwe94}.
But notice that a spacetime with helical symmetry and standing gravitational
waves cannot be asymptotically flat \cite{GibboS83}. 

To treat helically symmetric spacetimes, it is natural to choose coordinates
$(t,x^i)$ that are adapted to the symmetry, i.e. such that 
\be
	\wpar_t = \w{\ell} . 
\ee
Then all the fields are independent of the coordinate $t$. In particular,
\be \label{e:ini:dot_helic}
	\dot\tgm^{ij} = 0 \qquad \mbox{and} \qquad \dot K = 0 . 
\ee
If we employ the XCTS formalism to compute initial data, we therefore get some definite prescription for the free data $\dot\tgm^{ij}$ and $\dot K$. 
On the contrary, the requirements (\ref{e:ini:dot_helic}) do not have any immediate
translation in the CTT formalism. 

\begin{remark}
Helical symmetry can also be useful to treat binary black holes outside
the scope of the 3+1 formalism, as shown by Klein \cite{Klein04}, who
developed a quotient space formalism to reduce the problem to a 
three dimensional $\mathrm{SL}(2,\R)/\mathrm{SO}(1,1)$ sigma model.
\end{remark}

Taking into account (\ref{e:ini:dot_helic}) and choosing maximal slicing 
($K=0$), the XCTS system (\ref{e:ini:XCTS1})-(\ref{e:ini:XCTS3}) becomes
\bea
 \fl \tD_i \tD^i \Psi -\frac{{\tilde R}}{8}  \Psi
	+ \frac{1}{8} \hA_{ij} \hA^{ij} \, \Psi^{-7}
	+ 2\pi {\tilde E} \Psi^{-3}  = 0  \label{e:ini:XCTS-heli1} \\
 \fl  \tD_j \left( \frac{1}{\tilde N} (\tilde L \beta)^{ij} \right)
 - 16\pi {\tilde p}^i  = 0 \label{e:ini:XCTS-heli2} \\
 \fl \tD_i \tD^i (\tilde N \Psi^7) - (\tilde N \Psi^7)
	\left[ \frac{\tilde R}{8} 
	 + \frac{7}{8} \hA_{ij} \hA^{ij} \Psi^{-8} 
	+ 2 \pi (\tilde E+2\tilde S) \Psi^{-4} \right]  = 0  ,\label{e:ini:XCTS-heli3}
\eea
where [cf. Eq.~(\ref{e:ini:hA_dg_beta})]
\be 
   \hA^{ij} = \frac{1}{2\tilde N} 
	(\tilde L \beta)^{ij} . 
\ee

\subsection{Helical symmetry and IWM approximation}

If we choose, as part of the free data, the conformal metric to be flat,
\be \label{e:ini:helic_tgm_f}
	\tgm_{ij} = f_{ij} , 
\ee
then the helically symmetric XCTS system  
(\ref{e:ini:XCTS-heli1})-(\ref{e:ini:XCTS-heli3}) reduces to 
\bea
 & & \Delta \Psi 
	+ \frac{1}{8} \hA_{ij} \hA^{ij} \, \Psi^{-7}
	+ 2\pi {\tilde E} \Psi^{-3} = 0 \label{e:ini:XCTS-heli-flat1} \\
 & & \Delta \beta^i + \frac{1}{3} \Df^i \Df_j \beta^j
	- (L\beta)^{ij} \Df_j \ln \tilde N
  =  16\pi {\tilde N} {\tilde p}^i  \label{e:ini:XCTS-heli-flat2} \\
& & \Delta(\tilde N \Psi^7) - (\tilde N \Psi^7)
	\left[  \frac{7}{8} \hA_{ij} \hA^{ij} \Psi^{-8} 
	+ 2 \pi (\tilde E+2\tilde S) \Psi^{-4} \right]= 0 , 
					\label{e:ini:XCTS-heli-flat3}
\eea
where 
\be 
   \hA^{ij} = \frac{1}{2\tilde N}  (L \beta)^{ij}  
\ee
and $\wDf$ is the connection associated with the flat metric $\w{f}$, 
$\Delta :=  \Df_i \Df^i$ is the flat Laplacian [Eq.~(\ref{e:ini:def_Delta})], and 
$(L\beta)^{ij} := \Df^i \beta^j + \Df^j \beta^i - \frac{2}{3} \Df_k\beta^k \, f^{ij}$ 
[Eq.~(\ref{e:ini:conf_Killing_def}) with $\tD^i = \Df^i$]. 

We remark that the system (\ref{e:ini:XCTS-heli-flat1})-(\ref{e:ini:XCTS-heli-flat3})
is identical to the system defining the Isenberg-Wilson-Mathews approximation to
general relativity \cite{Isenb78,WilsoM89} 
(see e.g. Sec.~6.6 of Ref.~\cite{Gourg07a}). This means that, within helical symmetry, 
the XCTS system with the choice $K=0$
and $\wtgm = \w{f}$ is equivalent to the IWM system. 
\begin{remark}
Contrary to IWM, XCTS is not some approximation to general relativity: it 
provides exact initial data. The only thing that may be questioned is
the astrophysical relevance of the XCTS
data with $\wtgm = \w{f}$. 
\end{remark}

\subsection{Initial data for orbiting binary black holes}

The concept of helical symmetry for generating orbiting binary black hole
initial data has been introduced in 2002 by Gourgoulhon, Grandcl\'ement and
Bonazzola \cite{GourgGB02,GrandGB02}. The system of equations that 
these authors have derived is equivalent to the XCTS system with $\wtgm = \w{f}$, 
their work being previous to the formulation of the XCTS method by Pfeiffer 
and York (2003) \cite{PfeifY03}. Since then other groups have combined 
XCTS with helical symmetry to compute binary black hole
initial data \cite{CookP04,Ansor05,Ansor07,CaudiCGP06}. Since all these studies
are using a flat conformal metric [choice (\ref{e:ini:helic_tgm_f})], 
the PDE system to be solved is
(\ref{e:ini:XCTS-heli-flat1})-(\ref{e:ini:XCTS-heli-flat3}), with 
the additional simplification $\tilde E = 0$ and ${\tilde p}^i=0$
(vacuum). The initial data manifold $\Sigma_0$ is chosen to be 
$\R^3$ minus two balls:
\be \label{e:ini:S0_R3_balls}
	\Sigma_0 = \R^3 \backslash (\mathcal{B}_1 \cup \mathcal{B}_2 ) .
\ee
In addition to the asymptotic flatness conditions, some boundary conditions
must be provided on the surfaces $\Sp_1$ and $\Sp_2$ of 
$\mathcal{B}_1$ and $\mathcal{B}_2$. 
One choose boundary conditions corresponding to a \emph{non-expanding horizon},
since this concept
characterizes black holes in equilibrium. We shall not detail these boundary
conditions here; they can be found in Refs.~\cite{CookP04,Dain04,DainJK05,GourgJ06a,JaramAL07}. 
The condition of non-expanding horizon provides 3 among the 5 required boundary conditions [for the 5 components $(\Psi,\tilde N,\beta^i)$]. The two remaining 
boundary conditions are given by (i) the choice of the foliation (choice of the
value of $N$ at $\Sp_1$ and $\Sp_2$) and (ii) the choice of the rotation state 
of each black hole (``individual spin''), as explained in Ref.~\cite{CaudiCGP06}. 

Numerical codes for solving the above system have been constructed by
\begin{itemize}
\item  Grandcl\'ement, Gourgoulhon and Bonazzola (2002) \cite{GrandGB02}
for corotating binary black holes;
\item Cook, Pfeiffer, Caudill and Grigsby (2004, 2006) \cite{CookP04,CaudiCGP06}
for corotating and irrotational binary black holes;
\item  Ansorg (2005, 2007) \cite{Ansor05,Ansor07}
for corotating binary black holes.
\end{itemize}
Detailed comparisons with post-Newtonian initial data (either from the standard 
post-Newtonian formalism \cite{Blanc02} or from the Effective One-Body
approach \cite{BuonaD99,Damou01}) have revealed a very good agreement,
as shown in Refs.~\cite{DamouGG02,CaudiCGP06}. 

An alternative to (\ref{e:ini:S0_R3_balls}) for the initial data manifold
would be to consider the twice-punctured $\R^3$:
\be \label{e:ini:S0_R3_points}
	\Sigma_0 = \R^3 \backslash \{O_1,O_2\} ,
\ee
where $O_1$ and $O_2$ are two points of $\R^3$. 
This would constitute some extension to the two bodies case of the punctured initial
data discussed in Sec.~\ref{s:ini:XCTS_work}. 
However, as shown by Hannam, Evans, Cook and Baumgarte in 2003 \cite{HannaECB03},
it is not possible to find a solution of the helically symmetric XCTS system
with a regular lapse in this case\footnote{see however Ref.~\cite{Hanna05} for some
attempt to circumvent this}. 
For this reason, initial data based on the puncture manifold (\ref{e:ini:S0_R3_points})
are computed within the CTT framework discussed in Sec.~\ref{s:ini:CTT}.
As already mentioned, there is no natural way to implement helical symmetry in
this framework. One instead selects the free data $\hA^{ij}_{\rm TT}$ to vanish
identically, as in the single black hole case treated in Secs.~\ref{s:ini:cflat_static}
and \ref{s:ini:Bowen_York}. Then
\be
	\hA^{ij} = (\tilde L X)^{ij} .
\ee
The vector $\w{X}$ must obey Eq.~(\ref{e:ini:mom_ex1}), which arises from the
momentum constraint. Since this equation is linear, 
one may choose for $\w{X}$ a linear superposition of two Bowen-York
solutions (Sec.~\ref{s:ini:Bowen_York}):
\be
	\w{X} = \w{X}_{(\w{P}^{(1)},\w{S}^{(1)})}
	+ \w{X}_{(\w{P}^{(2)},\w{S}^{(2)})} ,
\ee
where $\w{X}_{(\w{P}^{(a)},\w{S}^{(a)})}$ ($a=1,2$) is the Bowen-York
solution (\ref{e:ini:BY_X}) centered on $O_a$.
This method has been first implemented by Baumgarte in 2000 \cite{Baumg00}. 
It has been since then used by Baker, Campanelli, Lousto and Takashi
(2002) \cite{BakerCLT02} and Ansorg, Br\"ugmann and Tichy (2004) \cite{AnsorBT04}.
The initial data hence obtained are closed from helically symmetric XCTS initial 
data at large separation but deviate significantly from them, as well as
from post-Newtonian initial data, when the two black holes are very close.
This means that the Bowen-York extrinsic curvature is bad for close binary systems
in quasi-equilibrium
(see discussion in Ref.~\cite{DamouGG02}).
\begin{remark}
Despite of this, CTT Bowen-York configurations have been used as initial data
for the recent binary black hole inspiral and merger computations 
by Baker et al. \cite{BakerCCKV06a,BakerCCKV06b,VanMeBKC06} and Campanelli et al. 
\cite{CampaLMZ06,CampaLZ06a,CampaLZ06b,CampaLZ06c}. Fortunately, these initial
data had a relative large separation, so that they differed only slightly
from the helically symmetric XCTS ones. 
\end{remark}

Instead of choosing somewhat arbitrarily the free data of the CTT and XCTS methods,
notably setting $\wtgm=\w{f}$, one may deduce them from post-Newtonian results. 
This has been done for the binary black hole problem by Tichy, Br\"ugmann,
Campanelli and Diener (2003) \cite{TichyBCD03}, who have used the CTT
method with the free data $(\tgm_{ij},\hA^{ij}_{\rm TT})$ given by the second order
post-Newtonian (2PN) metric. 
This work has been improved recently by Kelly, Tichy, Campanelli and Whiting (2007)
\cite{KellyTCW07}.
In the same spirit, Nissanke (2006) \cite{Nissa06}
has provided 2PN free data for both the CTT and XCTS methods. 

\subsection{Initial data for orbiting binary neutron stars}

For computing initial data corresponding to orbiting binary neutron stars,
one must solve equations for the fluid motion in addition to the Einstein constraints. 
Basically this amounts to 
solving $\nabla_\nu T^{\mu\nu} = 0$ 
in the context of helical symmetry. One can then show that a first integral
of motion exists in two cases: (i) the stars are corotating, i.e. the fluid 4-velocity
is colinear to the helical Killing vector (rigid motion), 
(ii) the stars are irrotational, i.e. 
the fluid vorticity vanishes. The most straightforward way to get the first 
integral of motion
is by means of the Carter-Lichnerowicz formulation of relativistic hydrodynamics,
as shown in Sec.~7 of Ref.~\cite{Gourg06}. Other derivations have been obtained
in 1998 by Teukolsky \cite{Teuko98} and Shibata \cite{Shiba98}. 

From the astrophysical point of view, the irrotational motion is much more interesting
than the corotating one, because the viscosity of neutron star matter is far too 
low to ensure the synchronization of the stellar spins with the orbital motion. 
On the other side, the irrotational state is a very good approximation for neutron
stars that are not millisecond rotators. Indeed, for these stars 
the spin frequency is much lower than the orbital frequency at 
the late stages of the inspiral and thus can be neglected. 

The first initial data for binary neutron stars on circular orbits have been computed
by Baumgarte, Cook, Scheel, Shapiro and Teukolsky in 1997 
\cite{BaumgCSST97,BaumgCSST98}
in the corotating case, and by Bonazzola, Gourgoulhon and Marck in 1999 \cite{BonazGM99a} in the irrotational case. These results were based on a polytropic equation of state.
Since then configurations in the irrotational regime have been obtained
\begin{itemize}
\item for a polytropic equation of state 
\cite{MarroMW99,UryuE00,UryuSE00,GourgGTMB01,TanigG02b,TanigG03} (the configurations
obtained in Ref.~\cite{TanigG03} have been used as initial data by Shibata \cite{Shiba07} to compute the merger of binary neutron stars);
\item for nuclear matter equations of state issued from recent nuclear physics
computations \cite{BejgeGGHTZ05,OechsJM07};
\item for strange quark matter \cite{OechsUPT04,LimouGG05}.
\end{itemize}
All these computation are based on a flat conformal metric [choice (\ref{e:ini:helic_tgm_f})], by solving the helically symmetric XCTS system
(\ref{e:ini:XCTS-heli-flat1})-(\ref{e:ini:XCTS-heli-flat3}), supplemented by
an elliptic equation for the velocity potential.
Only very recently, configurations based on a non flat conformal metric
have been obtained by Uryu, Limousin, Friedman, Gourgoulhon and Shibata
\cite{UryuLFGS06}. The conformal metric is then deduced from a waveless approximation
developed by Shibata, Uryu and Friedman \cite{ShibaUF04} and 
which goes beyond the IWM approximation.

\subsection{Initial data for black hole - neutron star binaries}

Let us mention briefly that initial data for a mixed binary system, i.e. 
a system composed of a black hole and a neutron star, have been obtained
very recently by Grandcl\'ement \cite{Grand06} and Taniguchi, Baumgarte, 
Faber and Shapiro \cite{TanigBFS06,TanigBFS07}.
Codes aiming at computing such systems have also been presented by 
Ansorg \cite{Ansor07} and Tsokaros and Uryu \cite{TsokaU07}.

\ack{I warmly thank the organizers of the  VII Mexican school, namely
Miguel Alcubierre, Hugo Garcia-Compean and Luis Urena, for their support and the
success of the school.  
I also express my gratitude to Marcelo Salgado for his help and many discussions
and to Nicolas Vasset for the careful reading of the manuscript.}


\section*{References}
\begin{thebibliography}{10}

\bibitem{Ansor05}
M. Ansorg : {\em  Double-domain spectral method for black hole excision data},
Phys. Rev. D {\bf 72}, 024018 (2005). 

\bibitem{Ansor07}
M. Ansorg: {\em Multi-Domain Spectral Method for Initial Data of Arbitrary Binaries in General Relativity}, 
Class. Quantum Grav. {\bf 24}, S1 (2007).

\bibitem{AnsorBT04}
M. Ansorg, B. Br\"ugmann and W. Tichy : {\em  Single-domain spectral method for black hole puncture data},
Phys. Rev. D {\bf 70}, 064011 (2004). 

\bibitem{BaierSW62}
R.F. Baierlein, D.H Sharp and J.A. Wheeler : {\em Three-Dimensional Geometry as
Carrier of Information about Time}, 
Phys. Rev. {\bf 126}, 1864 (1962).

\bibitem{BakerCLT02}
J.G. Baker, M. Campanelli, C.O. Lousto and R. Takahashi : 
{\em Modeling gravitational radiation from coalescing binary black holes},
Phys. Rev. D {\bf 65}, 124012 (2002). 

\bibitem{BakerCCKV06a}
J.G. Baker, J. Centrella, D.-I. Choi, M. Koppitz, and J. van Meter : 
{\em  Gravitational-Wave Extraction from an Inspiraling Configuration 
of Merging Black Holes},
Phys. Rev. Lett. {\bf 96}, 111102 (2006). 

\bibitem{BakerCCKV06b}
J.G. Baker, J. Centrella, D.-I. Choi, M. Koppitz, and J. van Meter : 
{\em  Binary black hole merger dynamics and waveforms},
Phys. Rev. D {\bf 73}, 104002 (2006). 

\bibitem{Bartn93}
R. Bartnik : {\em Quasi-spherical metrics and prescribed scalar curvature},
J. Diff. Geom. {\bf 37}, 31 (1993).

\bibitem{BartnF93}
R. Bartnik and G. Fodor : {\em On the restricted validity of the thin sandwich conjecture}, 
Phys. Rev. D {\bf 48}, 3596 (1993). 

\bibitem{BartnI04}
R. Bartnik and J. Isenberg : {\em The Constraint Equations},
in {\em The Einstein Equations and the Large Scale Behavior of 
Gravitational Fields --- 50 years of the Cauchy Problem in General Relativity}, 
edited by P.T.~Chru\'sciel and H.~Friedrich,
Birkh\"auser Verlag, Basel (2004), p.~1.

\bibitem{Baumg00}
T.W.~Baumgarte : {\em Innermost stable circular orbit of binary black holes},
Phys. Rev. D {\bf 62}, 024018 (2000). 

\bibitem{BaumgCSST97}
T.W. Baumgarte, G.B. Cook, M.A. Scheel, S.L. Shapiro, and S.A. Teukolsky :
{\em Binary neutron stars in general relativity: Quasiequilibrium models},
Phys. Rev. Lett. {\bf 79}, 1182 (1997).

\bibitem{BaumgCSST98}
T.W. Baumgarte, G.B. Cook, M.A. Scheel, S.L. Shapiro, and S.A. Teukolsky :
{\em General relativistic models of binary neutron stars in quasiequilibrium},
Phys. Rev. D {\bf 57}, 7299 (1998).

\bibitem{BaumgOP06}
T.W. Baumgarte, N. \'O Murchadha, and H.P. Pfeiffer :
{\em Einstein constraints: Uniqueness and non-uniqueness in the conformal 
thin sandwich approach},
Phys. Rev. D {\bf 75}, 044009 (2007).

\bibitem{BeigK04}
R. Beig and W. Krammer : {\em Bowen-York tensors},
Class. Quantum Grav. {\bf 21}, S73 (2004).  

\bibitem{BejgeGGHTZ05}
M. Bejger, D. Gondek-Rosi{\'n}ska, E. Gourgoulhon, P. Haensel, 
K. Taniguchi, and  J. L. Zdunik~: 
{\em Impact of the nuclear equation of state on 
the last orbits of binary neutron stars},
Astron. Astrophys. {\bf 431}, 297-306 (2005).

\bibitem{Blanc02}
L. Blanchet : {\em Innermost circular orbit of binary black holes at the third post-Newtonian approximation},
Phys. Rev. D {\bf 65}, 124009 (2002). 

\bibitem{Blanc06a}
L. Blanchet : {\em Gravitational Radiation from Post-Newtonian Sources and Inspiralling Compact Binaries}, 
Living Rev. Relativity {\bf 9}, 4 (2006);
\verb+http://www.livingreviews.org/lrr-2006-4+

\bibitem{BonazGM99a}
S.~Bonazzola, E.~Gourgoulhon, and J.-A. Marck :
{\em Numerical models of irrotational binary neutron stars in
general relativity},
Phys. Rev. Lett. {\bf 82}, 892 (1999).

\bibitem{BowenY80}
J.M. Bowen and J.W. York : {\em Time-asymmetric initial data for black holes and black-hole collisions},
Phys. Rev. D {\bf 21}, 2047 (1980).

\bibitem{BrandB97}
S. Brandt and B. Br\"ugmann : {\em A Simple Construction of Initial Data for 
Multiple Black Holes}, 
Phys. Rev. Lett. {\bf 78}, 3606 (1997). 

\bibitem{BrandS95b}
S.R. Brandt and E. Seidel : {\em Evolution of distorted rotating black holes. II. Dynamics and analysis}, 
Phys. Rev. D {\bf 52}, 870 (1995).

\bibitem{BuonaD99}
A. Buonanno and T. Damour : {\em Effective one-body approach to general relativistic two-body dynamics}, 
Phys. Rev. D {\bf 59}, 084006 (1999). 

\bibitem{Campa07}
M. Campanelli : {\em The dawn of a golden age for binary black hole simulations},
in these proceedings.

\bibitem{CampaLMZ06}
M. Campanelli, C. O. Lousto, P. Marronetti, and Y. Zlochower :
{\em  Accurate Evolutions of Orbiting Black-Hole Binaries without Excision}, 
Phys. Rev. Lett. {\bf 96}, 111101 (2006). 

\bibitem{CampaLZ06a}
M. Campanelli, C. O. Lousto, and Y. Zlochower : 
{\em  Last orbit of binary black holes},
Phys. Rev. D {\bf 73}, 061501(R) (2006). 

\bibitem{CampaLZ06b}
M. Campanelli, C. O. Lousto, and Y. Zlochower : 
{\em Spinning-black-hole binaries: The orbital hang-up},
 Phys. Rev. D {\bf 74}, 041501(R) (2006).

\bibitem{CampaLZ06c}
M. Campanelli, C. O. Lousto, and Y. Zlochower : 
{\em  Spin-orbit interactions in black-hole binaries},
Phys. Rev. D {\bf 74}, 084023 (2006). 

\bibitem{Canto77}
M. Cantor:  {\em The existence of non-trivial asymptotically flat initial data for vacuum spacetimes}, 
Commun. Math. Phys. {\bf 57}, 83 (1977). 

\bibitem{Canto79}
M. Cantor : {\em Some problems of global analysis on asymptotically 
simple manifolds},
Compositio Mathematica {\bf 38}, 3 (1979); 
available at \verb+http://www.numdam.org/item?id=CM_1979__38_1_3_0+

\bibitem{CaudiCGP06}
M. Caudill, G.B. Cook, J.D. Grigsby, and H.P. Pfeiffer : {\em  Circular orbits and spin in black-hole initial data}, 
Phys. Rev. D {\bf 74}, 064011 (2006). 

\bibitem{Chopt07}
M.W. Choptuik : {\em Numerical analysis for numerical relativists},
in these proceedings. 

\bibitem{Choqu71}
Y. Choquet-Bruhat : {\em New elliptic system and global solutions for the
constraints equations in general relativity}, 
Commun. Math. Phys. {\bf 21}, 211 (1971). 

\bibitem{ChoquC81}
Y. Choquet-Bruhat and D. Christodoulou : {\em Elliptic systems of $H_{s,\delta}$
spaces on manifolds which are Euclidean at infinity}, Acta Math. {\bf 146}, 129 (1981)

\bibitem{ChoquIY00}
Y. Choquet-Bruhat, J. Isenberg, and J.W. York : 
{\em Einstein constraints on asymptotically Euclidean manifolds},
Phys. Rev. D {\bf 61}, 084034 (2000). 

\bibitem{ChoquY80}
Y. Choquet-Bruhat and J.W. York : {\em The Cauchy Problem},
in {\em General Relativity and Gravitation, one hundred Years after the Birth
of Albert Einstein}, Vol.~1, edited by A. Held,
Plenum Press, New York (1980), p.~99.

\bibitem{Cook00} 
G.B. Cook : {\em Initial data for numerical relativity},
Living Rev. Relativity {\bf 3}, 5 (2000);
\verb+http://www.livingreviews.org/lrr-2000-5+

\bibitem{CookP04}
G.B. Cook and H.P. Pfeiffer : {\em  Excision boundary conditions for black-hole initial data}, 
Phys. Rev. D {\bf 70}, 104016 (2004). 

\bibitem{Corvi00}
J. Corvino : {\em Scalar curvature deformation and a gluing construction for the
Einstein constraint equations},
Commun. Math. Phys. {\bf 214}, 137 (2000). 

\bibitem{Dain04}
S. Dain : {\em Trapped surfaces as boundaries for the constraint equations},
Class. Quantum Grav. {\bf 21}, 555 (2004); 
errata in Class. Quantum Grav. {\bf 22}, 769 (2005). 

\bibitem{DainJK05}
S. Dain, J.L. Jaramillo, and B. Krishnan : {\em On the existence of initial data 
containing isolated black holes},
Phys.Rev. D {\bf 71}, 064003 (2005). 

\bibitem{Damou01}
T. Damour : {\em Coalescence of two spinning black holes: An effective one-body approach},
Phys. Rev. D {\bf 64}, 124013 (2001). 

\bibitem{DamouGG02}
T.~Damour, E.~Gourgoulhon, and P.~Grandcl\'ement :
{\em Circular orbits of corotating binary black holes: comparison between
analytical and numerical results},
Phys. Rev. D {\bf 66}, 024007 (2002).

\bibitem{Detwe94}
S. Detweiler : {\em Periodic solutions of the Einstein equations for binary systems}, 
Phys. Rev. D {\bf 50}, 4929 (1994).

\bibitem{Foure56}
Y. Four\`es-Bruhat (Y. Choquet-Bruhat) : {\em Sur l'Int\'egration des \'Equations de
la Relativit\'e G\'en\'erale}, 
J. Rational Mech. Anal. {\bf 5}, 951 (1956). 

\bibitem{FriedUS02}
J.L. Friedman, K. Uryu and M. Shibata : {\em Thermodynamics of binary black holes and neutron stars}, 
Phys. Rev. D {\bf 65}, 064035 (2002);
erratum in  Phys. Rev. D {\bf 70}, 129904(E) (2004).

\bibitem{GaratP00}
A. Garat and R.H. Price : {\em Nonexistence of conformally flat slices of the Kerr spacetime},
Phys. Rev. D {\bf 61}, 124011 (2000).

\bibitem{GibboS83}
G.W.~Gibbons and J.M.~Stewart : {\em Absence of asymptotically flat solutions
of Einstein's equations which are periodic and empty near infinity},
in {\em Classical General Relativity},
Eds.~W.B.~Bonnor, J.N.~Islam and M.A.H.~MacCallum
Cambridge University Press, Cambridge (1983), p.~77.

\bibitem{GleisNPP98}
R.J. Gleiser, C.O. Nicasio, R.H. Price, and J. Pullin : {\em Evolving the Bowen-York
initial data for spinning black holes},
Phys. Rev. D {\bf 57}, 3401 (1998).

\bibitem{Gourg06}
E. Gourgoulhon : {\em  An introduction to relativistic hydrodynamics},
in {\em Stellar Fluid Dynamics and Numerical Simulations: From the Sun
to Neutron Stars}, 
edited by M. Rieutord \& B. Dubrulle,
EAS Publications Series {\bf 21}, 
EDP Sciences, Les Ulis (2006), p.~43; 
available as arXiv:gr-qc/0603009.

\bibitem{Gourg07a}
E. Gourgoulhon : {\em 3+1 Formalism and Bases of Numerical Relativity}, 
lectures at Institut Henri Poincar\'e (Paris, Sept.-Dec. 2006), arXiv:gr-qc/0703035. 

\bibitem{GourgGB02}
E.~Gourgoulhon, P.~Grandcl\'ement, and S.~Bonazzola :
{\em Binary black holes in circular orbits. I. A global spacetime approach},
Phys. Rev. D {\bf 65}, 044020 (2002).

\bibitem{GourgGTMB01}
E.~Gourgoulhon, P.~Grandcl\'ement, K.~Taniguchi, J.-A.~Marck, and
S.~Bonazzola :
{\em Quasiequilibrium sequences of synchronized and irrotational binary neutron
stars in general relativity: Method and tests},
Phys. Rev. D {\bf 63}, 064029 (2001).

\bibitem{GourgJ06a} 
E. Gourgoulhon and J.L. Jaramillo :
{\em A 3+1 perspective on null hypersurfaces and isolated horizons},
Phys. Rep. {\bf 423}, 159 (2006).

\bibitem{Grand06}
P. Grandcl\'ement : {\em  Accurate and realistic initial data for black hole-neutron star binaries},
Phys. Rev. D {\bf 74}, 124002 (2006); erratum in Phys. Rev. D {\bf 75}, 129903(E) (2007).

\bibitem{GrandBGM01}
P.~Grandcl\'ement, S.~Bonazzola, E.~Gourgoulhon, and J.-A.~Marck :
{\em A multi-domain spectral method for scalar and vectorial Poisson
equations with non-compact sources},
J. Comput. Phys. {\bf 170}, 231 (2001).

\bibitem{GrandGB02}
P.~Grandcl\'ement, E.~Gourgoulhon, and S.~Bonazzola :
{\em Binary black holes in circular orbits.
II. Numerical methods and first results},
Phys. Rev. D {\bf 65}, 044021 (2002).

\bibitem{GrandN07}
P.~Grandcl\'ement and J. Novak : {\em Spectral methods for numerical relativity},
Living Rev. Relativity, submitted, preprint arXiv:0706.2286.

\bibitem{Hanna05}
M.D. Hannam : {\em Quasicircular orbits of conformal thin-sandwich puncture binary 
black holes}, 
Phys. Rev. D {\bf 72}, 044025 (2005).

\bibitem{HannaECB03}
M.D. Hannam, C.R. Evans, G.B Cook and T.W. Baumgarte : {\em Can a combination of 
the conformal thin-sandwich and puncture methods yield binary black hole
solutions in quasiequilibrium?},
Phys. Rev. D {\bf 68}, 064003 (2003).

\bibitem{Isenb78}
J.A.~Isenberg : {\em Waveless Approximation Theories of Gravity},
preprint University of Maryland (1978), unpublished but 
available as arXiv:gr-qc/0702113; 
an abridged version can be found in Ref.~\cite{IsenbN80}.

\bibitem{Isenb95}
J. Isenberg : {\em Constant mean curvature solutions of the Einstein
constraint equations on closed manifolds},
Class. Quantum Grav. {\bf 12}, 2249 (1995). 

\bibitem{IsenbMP02}
J. Isenberg, R. Mazzeo, and D. Pollack : {\em Gluing and wormholes for the Einstein
constraint equations},
Commun. Math. Phys. {\bf 231}, 529 (2002). 

\bibitem{IsenbN80}
J.~Isenberg and J.~Nester : {\em Canonical Gravity}, 
in {\em General Relativity and Gravitation, one hundred Years after the Birth
of Albert Einstein},
Vol.~1, edited by A.~Held,
Plenum Press, New York (1980), p.~23. 

\bibitem{JaramAL07}
J.L. Jaramillo, M. Ansorg, F. Limousin : {\em Numerical implementation of isolated horizon boundary conditions},
Phys. Rev. D {\bf 75}, 024019 (2007). 

\bibitem{KellyTCW07}
B.J. Kelly, W. Tichy, M. Campanelli, and B.F. Whiting : {\em  Black-hole puncture initial data with realistic gravitational wave content}, 
Phys. Rev. D {\bf 76}, 024008 (2007).

\bibitem{Klein04}
C. Klein : {\em  Binary black hole spacetimes with a helical Killing vector},
Phys. Rev. D {\bf 70}, 124026 (2004). 

\bibitem{Lagun04}
P. Laguna : {\em Conformal-thin-sandwich initial data for a single boosted or 
spinning black hole puncture}, 
Phys. Rev. D {\bf 69}, 104020 (2004). 

\bibitem{Lagun07}
P. Laguna : {\em Two and three body encounters: Astrophysics and the role of numerical 
relativity}, in these proceedings. 

\bibitem{Lichn44}
A. Lichnerowicz : {\em L'int\'egration des \'equations de la gravitation 
relativiste et le probl\`eme des n corps}, 
J. Math. Pures Appl. {\bf 23}, 37 (1944); reprinted in 
A. Lichnerowicz : {\em Choix d'\oe uvres math\'ematiques}, 
Hermann, Paris (1982), p.~4.

\bibitem{Lichn52}
A. Lichnerowicz : {\em Sur les \'equations relativistes de la gravitation},
Bulletin de la S.M.F. {\bf 80}, 237 (1952);
available at \verb+http://www.numdam.org/item?id=BSMF_1952__80__237_0+

\bibitem{LimouGG05}
F. Limousin, D. Gondek-Rosi{\'n}ska, and E. Gourgoulhon : 
{\em Last orbits of binary strange quark stars},
Phys. Rev. D {\bf 71}, 064012 (2005).

\bibitem{MarroMW99}
P.~Marronetti, G.J. Mathews, and J.R. Wilson : {\em Irrotational
binary neutron stars in quasiequilibrium},
Phys. Rev. D {\bf 60}, 087301 (1999).

\bibitem{Maxwe04b}
D. Maxwell : {\em Initial Data for Black Holes and Rough Spacetimes},
PhD Thesis, University of Washington (2004). 

\bibitem{Nissa06}
S. Nissanke : {\em  Post-Newtonian freely specifiable initial data for binary 
black holes in numerical relativity},
Phys. Rev. D {\bf 73}, 124002 (2006).

\bibitem{OMurcY74}
N. \'O Murchadha and J.W. York : {\em Initial-value problem of general relativity. I. General
formulation and physical interpretation},
Phys. Rev. D {\bf 10}, 428 (1974). 

\bibitem{OechsJM07}
R. Oechslin, H.-T. Janka and A. Marek : {\em Relativistic neutron star merger simulations with non-zero temperature equations of state I. Variation of binary parameters and equation of state}, 
Astron. Astrophys. {\bf 467}, 395 (2007).

\bibitem{OechsUPT04}
R. Oechslin, K. Uryu, G. Poghosyan, and  F. K. Thielemann :
{\em The Influence of Quark Matter at High Densities on Binary Neutron Star Mergers},
Mon. Not. Roy. Astron. Soc. {\bf 349}, 1469 (2004). 

\bibitem{Pfeif04}
H.P. Pfeiffer : {\em The initial value problem in numerical relativity},
in {\em Proceedings Miami Waves Conference 2004}
[preprint arXiv:gr-qc/0412002].

\bibitem{PfeifY03}
H.P. Pfeiffer and J.W. York : {\em Extrinsic curvature and the Einstein
constraints},
Phys. Rev. D {\bf 67}, 044022 (2003). 

\bibitem{PfeifY05}
H.P. Pfeiffer and J.W. York : {\em Uniqueness and Nonuniqueness in the 
Einstein Constraints},
Phys. Rev. Lett. {\bf 95}, 091101 (2005).

\bibitem{Preto05b}
F. Pretorius : {\em  Evolution of Binary Black-Hole Spacetimes},
Phys. Rev. Lett. {\bf 95}, 121101 (2005).

\bibitem{Shiba98}
M. Shibata : {\em Relativistic formalism for computation of irrotational
binary stars in quasiequilibrium states},
Phys. Rev. D {\bf 58}, 024012 (1998).

\bibitem{Shiba07}
M. Shibata : {\em Merger of binary neutron stars in full general relativity},
in these proceedings. 

\bibitem{ShibaUF04}
M. Shibata, K. Uryu, and J.L. Friedman : {\em Deriving formulations for numerical computation of binary neutron stars in quasicircular orbits},
Phys. Rev. D {\bf 70}, 044044 (2004); errata in
Phys. Rev. D {\bf 70}, 129901(E) (2004).

\bibitem{Shoem07}
D. Shoemaker : {\em Binary Black Hole Simulations Through the Eyepiece of Data Analysis},
in these proceedings. 

\bibitem{SmarrY78a}
L.~Smarr and J.W.~York : {\em Radiation gauge in general relativity},
Phys. Rev. D {\bf 17}, 1945 (1978).

\bibitem{TanigBFS06}
K. Taniguchi, T.W. Baumgarte, J.A. Faber, and  S.L. Shapiro : 
{\em Quasiequilibrium sequences of black-hole-neutron-star binaries in general relativity}, 
Phys. Rev. D {\bf 74}, 041502(R) (2006). 

\bibitem{TanigBFS07}
K. Taniguchi, T.W. Baumgarte, J.A. Faber, and  S.L. Shapiro : 
{\em Quasiequilibrium black hole-neutron star binaries in general relativity},
Phys. Rev. D {\bf 75}, 084005 (2007).

\bibitem{TanigG02b}
K.~Taniguchi and E.~Gourgoulhon :
{\em Quasiequilibrium sequences of synchronized and irrotational binary neutron stars
in general relativity. III. Identical and different mass stars with $\gamma=2$},
Phys. Rev. D {\bf 66}, 104019 (2002).

\bibitem{TanigG03} 
K. Taniguchi and E. Gourgoulhon :
{\em Various features of quasiequilibrium sequences of
  binary neutron stars in general relativity},
Phys. Rev. D  {\bf 68}, 124025 (2003).

\bibitem{Teuko82}
S.A. Teukolsky : \emph{Linearized quadrupole waves in general relativity 
and the motion of test particles},
Phys. Rev. D {\bf 26}, 745 (1982). 

\bibitem{Teuko98}
S.A Teukolsky : {\em Irrotational binary neutron stars in
quasi-equilibrium in general relativity},
Astrophys. J. {\bf 504}, 442 (1998).

\bibitem{TichyBCD03}
W. Tichy, B. Br\"ugmann, M. Campanelli, and P. Diener : {\em Binary black hole initial data for numerical general relativity based on post-Newtonian data},
Phys. Rev. D {\bf 67}, 064008 (2003).

\bibitem{TsokaU07}
A.A. Tsokaros and K. Uryu : {\em  Numerical method for binary black hole/neutron star initial data: Code test},
Phys. Rev. D {\bf 75}, 044026 (2007). 

\bibitem{UryuE00}
K. Uryu and Y. Eriguchi : {\em New numerical method for
constructing quasiequilibrium sequences of irrotational
binary neutron stars in general relativity},
Phys. Rev. D {\bf 61}, 124023 (2000).

\bibitem{UryuSE00}
K. Uryu, M.~Shibata, and Y. Eriguchi : {\em Properties of general
relativistic, irrotational binary neutron stars in close
quasiequilibrium orbits: Polytropic equations of state},
Phys. Rev. D {\bf 62}, 104015 (2000).

\bibitem{UryuLFGS06}
K. Uryu, F. Limousin, J.L. Friedman, E. Gourgoulhon, and M. Shibata : 
{\em  Binary Neutron Stars: Equilibrium Models beyond Spatial Conformal Flatness},
Phys. Rev. Lett. {\bf 97}, 171101 (2006).

\bibitem{VanMeBKC06}
J.R. van Meter, J.G. Baker, M. Koppitz, D.I. Choi :
{\em How to move a black hole without excision: gauge conditions for the numerical evolution of a moving puncture},
Phys. Rev. D {\bf 73}, 124011 (2006).

\bibitem{Walsh06}
D. Walsh : {\em Non-uniqueness in conformal formulations of the
Einstein Constraints},
Class. Quantum Grav {\bf 24}, 1911 (2007). 

\bibitem{Wheel64}
J.A. Wheeler : {\em Geometrodynamics and the issue of the final state},
in {\em Relativity, Groups and Topology}, edited by
C. DeWitt and B.S. DeWitt, 
Gordon and Breach, New York (1964), p.~316.

\bibitem{WilsoM89}
J.R. Wilson and G.J. Mathews : {\em Relativistic hydrodynamics},
in {\em Frontiers in numerical relativity},
edited by C.R.~Evans, L.S.~Finn and D.W.~Hobill,
Cambridge University Press, Cambridge (1989), p.~306.

\bibitem{York72b}
J.W.~York : {\em Mapping onto Solutions of the Gravitational Initial Value Problem},
J. Math. Phys. {\bf 13}, 125 (1972). 

\bibitem{York73}
J.W.~York : {\em Conformally invariant orthogonal decomposition of symmetric 
tensors on Riemannian manifolds and the initial-value problem of general relativity},
J. Math. Phys. {\bf 14}, 456 (1973). 

\bibitem{York74}
J.W.~York : {\em Covariant decompositions of symmetric tensors in the 
theory of gravitation}, 
Ann. Inst. Henri Poincar\'e A {\bf 21}, 319 (1974); \\
available at \verb+http://www.numdam.org/item?id=AIHPA_1974__21_4_319_0+

\bibitem{York79}
J.W.~York : {\em Kinematics and dynamics of general relativity},
in {\em Sources of Gravitational Radiation}, edited by
L.L.~Smarr, Cambridge University Press, Cambridge (1979), p.~83.

\bibitem{York99}
J.W. York : {\em Conformal ``thin-sandwich'' data for the initial-value 
problem of general relativity},
Phys. Rev. Lett. {\bf 82}, 1350 (1999).

\bibitem{York04}
J.W. York : {\em Velocities and Momenta in an Extended Elliptic Form of the Initial Value Conditions},
Nuovo Cim. {\bf B119}, 823 (2004). 

\end{thebibliography}
\end{document}